\documentclass[preprint,amsmath,amssymb,aps,a4]{revtex4-2}
\usepackage{graphicx}
\usepackage{lineno}
\usepackage{bm}
\usepackage{txfonts}
\usepackage{hyperref}
\usepackage{caption}
\usepackage{amsmath}
\usepackage{amssymb}
\usepackage[section]{placeins}
\usepackage{tabularx}
\usepackage{braket}
\usepackage{multirow}
\usepackage{subcaption}
\usepackage{rotating}
\usepackage{xcolor}
\date{\today}
\newcommand{\be}{\begin{eqnarray}}
\newcommand{\ee}{\end{eqnarray}}

\newcommand{\bfk}{{\bf k}_{\perp}}

\begin{document}
\title{Scalar, Vector and Tensor Form Factors of Pion and Kaon}
\author{Satyajit Puhan$^{1}$}
\email{puhansatyajit@gmail.com}
  \author{Harleen Dahiya$^{1}$}
\email{dahiyah@nitj.ac.in}
\affiliation{$^1$ Computational High Energy Physics Lab, Department of Physics, Dr. B.R. Ambedkar National
	Institute of Technology, Jalandhar, 144008, India}

\date{\today}%
\begin{abstract}
We calculate the possible scalar, vector, and tensor form factors (FFs) for pion and kaon in the light-cone quark model (LCQM). These FFs are calculated from the leading and sub-leading twist generalized parton distribution functions (GPDs) of the pion and kaon. The vector and tensor FFs correspond to twist-2 GPDs, whereas the scalar FFs correspond to twist-3 scalar GPDs. We calculate the FFs from the GPDs quark-quark correlator and express them in the form of light-front wave functions (LFWFs). The behavior of these FFs was found to be in sync with experimental data, other model predictions, and lattice simulation results. We have also calculated the charge radii corresponding to different FFs and compared them with other models, lattice simulation results, and experimental data. The scalar radii of the pion and kaon of our model are found to be $0.528$ and $0.409$ $fm$, respectively.

\vspace{0.1cm}
    \noindent{\it Keywords}: Pion and kaon GPDs; scalar, vector and tensor form factors; pion and kaon form factors.
\end{abstract}
%
\maketitle
%
%

\section{Introduction\label{secintro}}
Being the lightest bound states in the hadron spectrum, the pion and kaon mesons are essential to comprehend the strong interaction in the non-perturbative domain of quantum chromodynamics (QCD) \cite{Brodsky:1997de, Zhang:1997dd}. They are pseudo-Goldstone bosons that originate from the spontaneous breaking of chiral symmetry and provide a novel perspective on the dynamical mass creation and confinement mechanisms in QCD \cite{Cui:2020dlm,Roberts:2021nhw}. Moreover, pions and kaons play a crucial role in explaining the long-range dynamics of the strong interactions. Understanding the internal structure of these hadrons is essential to provide a deeper insight into them and can be studied using multi-dimensional distribution functions like generalized transverse momentum parton distribution functions (GTMDs) \cite{Puhan:2025kzz,Zhang:2020ecj,Zhang:2021tnr}, generalized parton distributions (GPDs) \cite{Diehl:2003ny,Chavez:2021llq,Zhang:2021tnr,Broniowski:2022iip,Kaur:2018ewq,Guidal:2004nd}, transverse momentum parton distribution functions (TMDs) \cite{Diehl:2015uka,Angeles-Martinez:2015sea,Pasquini:2008ax,Kaur:2019jfa,Puhan:2023hio,Sharma:2024lal}, parton distribution functions (PDFs) \cite{Collins:1981uw, Martin:1998sq,Gluck:1994uf}, electromagnetic form factors (EMFFs) \cite{Xu:2023izo,Bijnens:2002hp,Acharyya:2024tql}, etc. GTMDs are called the mother distributions as they carry all the information about each degree of freedom of the hadron. The three-dimensional GPDs and TMDs carry information about the spatial and transverse structure along with the longitudinal momentum fraction of the valence quark inside a hadron. The quark GPDs are functions of longitudinal momentum fraction ($x$), skewness parameter ($\xi$), and transverse momentum transferred $\Delta_\perp^2$, while the TMDs are the functions of $x$ and transverse momentum of quark $\bfk$. The one-dimensional PDFs carry information about the longitudinal momentum fraction ($x$) carried by the valence quark from the hadron momenta \cite{Collins:1981uw,Martin:1998sq,Gluck:1994uf}. Therefore, to understand a complete picture of the internal structure of a hadron, one needs to study all these distribution functions non-perturbatively using low-energy models. These distributions can further be evolved to a high energy scale to compare with experimental observables with a proper evolution toolkit like Dokshitzer–Gribov–Lipatov–Altarelli–Parisi (DGLAP), Collins-Soper-Sterman (CSS), Balitsky-Fadin-Kuraev-Lipatov (BFKL) formalisms, etc \cite{Wang:2016sfq,Bautista:2016xnp,Aybat:2011ge}.
\par In this work, we have mainly focused on the form factors (FFs) of pion and kaon extracted from GPDs. There are a total of eight GPDs for pseudoscalar mesons up to twist-$4$ \cite{Meissner:2008ay}. The vector and tensor FFs are the result of leading twist unpolarized $F_1(x,\xi,-\Delta_\perp^2)$ and transversely polarized $H_1(x,\xi,-\Delta_\perp^2)$ GPDs respectively, while the scalar FF is derived from the scalar $E_2(x,\xi,-\Delta_\perp^2)$ GPD, which is twist-3 in nature. The coupling of a photon to the meson is described by the the widely studied EMFFs, which are the pion and kaon vector FFs and can be accessed directly through experiments \cite{NA7:1986vav,Bebek:1977pe,Barkov:1985ac,Quenzer:1978qt,DM2:1988xqd,Amendolia:1983di,Dally:1982zk,JeffersonLabFpi:2000nlc,JeffersonLabFpi-2:2006ysh,JeffersonLabFpi:2007vir,Dally:1980dj,Amendolia:1986ui,Carmignotto:2018uqj}. However, there is still a lack of information about this FF at a high $Q^2$ region. The pion vector FF has been investigated through lattice
QCD simulations \cite{Meyer:2011um,Frezzotti:2008dr,Boyle:2008yd,Meyer:2011um}, light-front holographic model (LFHM) \cite{Brodsky:2014yha}, AdS/QCD \cite{Kaur:2018ewq}, light-cone quark model (LCQM) \cite{Puhan:2024jaw}, Nambu–Jona-Lasinio model (NJL) \cite{Ninomiya:2014kja}, contact interaction \cite{Gutierrez-Guerrero:2010waf}, etc. The kaon vector FF has been explored through experiments \cite{Dally:1980dj,Amendolia:1986ui}, lattice simulations \cite{Kaneko:2010ru,Alexandrou:2021ztx} and in different theoretical models \cite{Abidin:2019xwu,Miramontes:2022uyi,Puhan:2024jaw}. The upcoming electron-ion collider (EIC) and upgraded 22 GeV JLAB E12-09-001 experiments aim to provide more insight into the vector FFs of pions and kaons \cite{AbdulKhalek:2021gbh,Arrington:2021biu}. The EIC is going to study the vector FFs of pion and kaon through the Sullivan process \cite{Accardi:2012qut}. In contrast, scalar and tensor FFs have remained relatively unexplored due to their lack of direct experimental accessibility and have not received sufficient attention. Similarly, very limited work has been reported for tensor and scalar FFs in theoretical models \cite{Wang:2022mrh}.
Notwithstanding, these two FFs still deserve
scrutiny. The scalar FF of the pion plays a significant role in hadron and nuclear physics, particularly due to its connection with low-energy $\pi \pi$ scattering processes. It is especially relevant in the isoscalar S-wave channel, where it provides insight into the dynamics of chiral symmetry breaking \cite{Dubnicka:2016bhn,Wang:2022mrh}. The scalar FF can be used to extract the pion's scalar radius and to constrain low-energy constants within chiral perturbation theory. While the study of tensor FFs not only provides deeper insights into the pion and kaon internal dynamics, it also  provides information about the
dipole-like distortion of the quark density in the transverse plane and can be used to study the gravitational FFs \cite{Freese:2019bhb}.

\par In this work, we have solved the quark-quark correlator for unpolarized and polarized valence quark GPDs to calculate the respective FFs using the LCQM. As we are mainly focusing on the comparative analysis of valence quarks, we have not considered the gluon in the present work. We have presented the FFs in the overlap form of light-front wave functions (LFWFs) and also in the explicit form by using the total wave function (spin and momentum space wave function). For a comprehensive understanding, we have plotted the FFs of pion, kaon, and their constituents as a function of $Q^2$ along with a comparison with available experimental data, lattice simulation data, and theoretical predictions. We have also evolved the FFs to high $Q^2$ using next-to-leading order (NLO) evolutions \cite{Son:2015bwa}. The vector, tensor, and scalar radii have also been calculated from the respective FFs and compared with available data for both the pion and kaon. For this work, we have calculated the FFs using the minimal Fock-state of LCQM \cite{Xiao:2002iv,Xiao:2003wf}. Being gauge-invariant and relativistic by nature, LCQM is a non-perturbative method.  Its primary focus is on valence quarks, which are the essential building blocks that determine the general structure and inherent characteristics of hadrons.

\par The paper is arranged as follows. In Sec. \ref{satya}, we have discussed the LCQM along with the spin and momentum wave function in LF formalism. The relations among the FFs and GPDs along with the GPDs quark-quark correlator, have been discussed in Sec. \ref{emff}. The overlap form with explicit expressions for all FFs have been presented in this section. While in Sec. \ref{result}, we have presented the plots for each FFs. The radii correspond to each FFs has also been calculated in this section. We have finally summarized our work in Sec. \ref{con}.

 \section{Methodology}\label{satya}
 \subsection{Light-cone quark model}\label{satya}
 The light-cone (LC) formalism provides a helpful framework for relativistically characterizing the hadrons in terms of quark and gluon degrees of freedom. The mesonic wave function can be expressed using the LC Fock-state expansion as \cite{Lepage:1980fj,Brodsky:1997de,Pasquini:2023aaf}
 \begin{eqnarray}
|M\rangle &=& \sum |q\bar{q}\rangle \psi_{q\bar{q}}
        + \sum
        |q\bar{q}g\rangle \psi_{q\bar{q}g} + \sum
        |q\bar{q}gg\rangle \psi_{q\bar{q}g g} + \cdots  \, ,
\end{eqnarray}
where $|M\rangle$ denotes the meson eigenstate. For this work, we have considered only the leading-order eigenstates of mesons having a quark-antiquark pair. However, the higher Fock-state contribution will provide a complete understanding of the gluons and sea-quarks inside a meson \cite{Pasquini:2023aaf}.
 The hadron wave function based on the  LC quantization of QCD  using multi-particle Fock-state expansion can be expressed as \cite{Puhan2023,Qian:2008px,Brodsky:2000xy}
\begin{eqnarray}\label{fockstate}
|M (P, S_z) \rangle
   &=&\sum_{n,\lambda_i}\int\prod_{i=1}^n \frac{\mathrm{d} x_i \mathrm{d}^2
        \mathbf{k}_{\perp i}}{\sqrt{x_i}}
 ~ \delta\Big(1-\sum_{i=1}^n x_i\Big)\delta^{(2)}\Big(\sum_{i=1}^n \mathbf{k}_{\perp i}\Big) ~\psi_{n/M}(x_i,\mathbf{k}_{\perp i},\lambda_i)   | n ; \mathbf{k}^+_i, \mathbf{k}_{\perp i},
        \lambda_i \rangle.
\end{eqnarray}
Here, $P=(P^+,P^-,P_{\perp})$ and $S_z$ are the four vector average momenta, and the longitudinal spin projection of the hadron having eigenstate $|M (P, S_z) \rangle$. $\mathbf{k_i}=(\mathbf{k}^+_i,\mathbf{k}^-_i,\mathbf{k}_{i \perp})$ is four vector momenta of the $i$-th constituent momentum of the hadron with helicity $\lambda_i$. $x_i=\mathbf{k}^+_i/P^+$ and $\mathbf{k}_{\perp i}$ are the longitudinal momentum fraction and transverse momenta of the $i$-th constituent of hadron. The total longitudinal and transverse momentum carried by the constituents obey the sum rule $\sum_{i=1}^n x_i=1$ and $\sum_{i=1}^n\textbf{k}_{\perp i}=0$.
\par As we are dealing with lower Fock-state calculations of pseudoscalar mesons pion and kaon, we have taken the minimal state description of Eq. (\ref{fockstate}) in the form of quark-antiquark which can be expressed as
\begin{eqnarray}
|\pi^+ (K^+)(P, S_Z=0)\rangle &=& \sum_{\lambda_q,\lambda_{\bar q}}\int
\frac{\mathrm{d} x \mathrm{d}^2
        \mathbf{k}_{\perp}}{16\pi^3\sqrt{x(1-x)}}
           \Psi_{\pi (K)}(x,\mathbf{k}_{\perp},\lambda_q,\lambda_{\bar q})|x P^+,\mathbf{k}_{\perp},
        \lambda_q,\lambda_{\bar q} \rangle
        .
        \label{meson}
\end{eqnarray}
Here, $|\pi^+ (K^+)(P, S_Z=0)\rangle$ is the eigenstate of pion (kaon). $\lambda_{q(\bar q)}$ is the helicity of quark (antiquark). $x$ and $1-x$ are the longitudinal momentum fractions of quark and antiquark, respectively. The four-vector average momenta of the meson ($P$), constituent quark ($k_q$) and antiquark ($k_{\bar q}$) in the LC frame are respectively defined as 
\begin{eqnarray}
P&\equiv&\bigg(P^+,\frac{\mathcal{M}^2_{\pi(K)}}{P^+},\textbf{0}_\perp \bigg)\label{n1},\\
k_{q}&\equiv&\bigg(x P^+, \frac{\textbf{k}_\perp^2+m_q^2}{x P^+},\textbf{k}_\perp \bigg),\\
k_{\bar q}&\equiv&\bigg((1-x) P^+, \frac{\textbf{k}_\perp^2+m_{\bar q}^2}{(1-x) P^+},-\textbf{k}_\perp \bigg),
\label{n3}
\end{eqnarray}
with $\mathcal{M}_{\pi(K)}$ being the invariant mass of the pion (kaon) defined in terms of its quark mass $m_q$ and antiquark mass $m_{\bar q}$ as
\begin{eqnarray}
    \mathcal{M}^2_{\pi(K)}=\frac{\bfk^2+m^2_q}{x} +\frac{\bfk^2+m^2_{\bar q}}{1-x}\, .
\end{eqnarray}
 $\Psi_{S_Z}(x,\mathbf{k}_{\perp},\lambda_i,\lambda_j)$ in Eq. (\ref{meson}) is the LC total wave function, which can be expanded in terms of momentum space wave function and spin wave function with different spin and helicity projections can be expressed as 
\begin{eqnarray}
\Psi_{\pi(K)}(x,\textbf{k}_\perp, \lambda_q, \lambda_{\bar q})= J_{S_z}(x,\textbf{k}_\perp, \lambda_q, \lambda_{\bar q}) \psi_{\pi(K)}(x, \textbf{k}_\perp).\
\label{space}
\end{eqnarray}
Here, $J_{S_z}(x,\textbf{k}_\perp, \lambda_q, \lambda_{\bar q})$ and $\psi_{\pi(K)}(x, \textbf{k}_\perp)$ are the spin and momentum space wave functions of the mesons respectively.
The momentum space wave function in Eq. (\ref{space}) can be expressed using quark masses and quark momentum through Brodsky-Huang-Lepage
(BHL) \cite{Qian:2008px,Xiao:2002iv} as 
\begin{eqnarray}
\psi_{\pi(K)}(x,\textbf{k}_\perp)= A_{\pi(K)} \ {\rm exp} \Bigg[-\frac{\frac{\textbf{k}^2_\perp+m_q^2}{x}+\frac{\textbf{k}^2_\perp+m^2_{\bar q}}{1-x}}{8 \beta_{\pi(K)}^2}
-\frac{(m_q^2-m_{\bar q}^2)^2}{8 \beta_{\pi(K)}^2 \bigg(\frac{\textbf{k}^2_\perp+m_q^2}{x}+\frac{\textbf{k}^2_\perp+m_{\bar q}^2}{1-x}\bigg)}+\frac{(m_q^2+m_{\bar q}^2)}{4 \beta_{\pi(K)}^2}\Bigg]\, ,
\label{bhl-k}
\end{eqnarray}
where $A_{\pi(K)}$ and $\beta_{\pi(K)}$ are the normalization constant and harmonic scale parameter of the pion (kaon), respectively. The normalization constant can be calculated through the normalizing condition as 
\begin{eqnarray}
    \int \frac{dx d^2\bfk}{2 (2\pi)^3}|\psi_{\pi(K)}(x,\bfk)|^2=1.
\end{eqnarray}
$J_{S_z}(x,\bfk,\lambda_q,\lambda_{\bar q})$ in Eq. (\ref{bhl-k}) is the front-form spin wave function derived from the instant form by Melosh-Wigner rotation \cite{Qian:2008px,Xiao:2002iv,Kaur:2020vkq}.
This transformation of instant-form state $\Phi(T)$ and front-form state $\Phi(F)$ is expressed as 
\begin{eqnarray}
\Phi_i^\uparrow(T)&=&-\frac{[\textbf{k}_i^R \Phi_i^\downarrow(F)-(\textbf{k}_i^+ +m_{q(\bar q)})\Phi_i^\uparrow(F)]}{\omega_i},\label{instant-front1}\\
\Phi_i^\downarrow(T)&=&\frac{[\textbf{k}_i^L\Phi_i^\uparrow(F)+(\textbf{k}_i^+ +m_{q(\bar q)})\Phi_i^\downarrow(F)]}{\omega_i}.
\label{instant-front}
\end{eqnarray}
Here, $\Phi(F)$ is a two-component Dirac spinor and $\textbf{k}_i^{R(L)}=\textbf{k}_i^1 \pm \iota \textbf{k}_i^2$. $\omega_i$ is defined as $\omega_i=1/ \sqrt{2 \textbf{k}^+_i (\textbf{k}^0+m_{q(\bar q)})}$. Now applying different momenta forms from Eqs. (\ref{n1})-(\ref{n3}) in  Melosh-Wigner rotation, the spin wave function is obtained in the form $\kappa_{0}^F(x,\textbf{k}_\perp, \lambda_q, \lambda_{\bar q})$ coefficient as 
\begin{eqnarray}
J_{S_z}(x,\textbf{k}_\perp, \lambda_q, \lambda_{\bar q})=\sum_{\lambda_q, \lambda_{\bar q}}\kappa_{0}^F(x,\textbf{k}_\perp, \lambda_q, \lambda_{\bar q}) \Phi_1^{\lambda_q}(F) \Phi_2^{\lambda_{\bar q}}(F).
\end{eqnarray}
These spin-wave function coefficients satisfy the following normalization relation
\begin{eqnarray}
\sum_{\lambda_q,\lambda_{\bar q}} \kappa_0^{F*}(x, \textbf{k}_\perp, \lambda_q, \lambda_{\bar q})\kappa_0^F(x, \textbf{k}_\perp, \lambda_q, \lambda_{\bar q})=1.
\end{eqnarray}
Similarly, the same spin-wave function can be calculated using the proper vertex chosen for the meson \cite{Choi:1996mq,Qian:2008px} as
\begin{eqnarray}
    J_{S_z}(x,\textbf{k}_\perp, \lambda_q, \lambda_{\bar q}) = \bar u (k_q,\lambda_q) \frac{\gamma_5}{\sqrt{2}\sqrt{\mathcal{M}_{\pi (K)}^2-(m^2_q-m^2_{\bar q}})} v(k_{\bar q},\lambda_{\bar q)} \, .
\end{eqnarray}
Here, $u$ and $v$ are the Dirac spinors.
Both the above methods give rise to the same form of spin wave function. The spin wave function for pseudoscalar mesons ($S_z=0$) with different helicities is expressed as \cite{Qian:2008px}
\begin{equation}
\left\{
  \begin{array}{lll}
    J_{(S_z=0)}(x,\mathbf{k}_\perp,\uparrow,\uparrow)&=&\frac{1}{\sqrt{2}}\omega_{\pi(K)}^{-1}(-\textbf{k}_{1\perp
    }+\iota \textbf{k}_{2\perp})(\mathcal{M}_{\pi(K)}+m_q+m_{\bar q}),\\
    J_{(S_z=0)}(x,\mathbf{k}_\perp,\uparrow,\downarrow)&=&\frac{1}{\sqrt{2}}\omega_{\pi(K)}^{-1}((1-x)m_q+x m_{\bar q})(\mathcal{M}_{\pi(K)}+m_q+m_{\bar q}),\\
    J_{(S_z=0)}(x,\mathbf{k}_\perp,\downarrow,\uparrow)&=&\frac{1}{\sqrt{2}}\omega_{\pi(K)}^{-1}(-(1-x)m_q-x m_{\bar q})(\mathcal{M}_{\pi(K)}+m_q+m_{\bar q}),\\
    J_{(S_z=0)}(x,\mathbf{k}_\perp,\downarrow,\downarrow)&=&\frac{-1}{\sqrt{2}}\omega_{\pi(K)}^{-1}(\textbf{k}_{1\perp}+\iota \textbf{k}_{2\perp})(\mathcal{M}_{\pi(K)}+m_q+m_{\bar q}),
  \end{array}
\right.
\end{equation}
with
$\omega_{\pi(K)}=(\mathcal{M}_{\pi(K)}+m_q+m_{\bar q})\sqrt{x(1-x)[\mathcal{M}_{\pi(K)}^2-(m_q-m_{\bar q})^2]}$. The two-particle Fock-state in Eq. (\ref{meson}) can be written in the form of LC wave functions (LCWFs) with all possible helicities of its constituent quark and antiquark as
\begin{eqnarray}
\ket{\pi^+(K^+) (P,S_z=0)}&=&\int \frac{{ {\rm d}x  \rm d}^2\textbf{k}_\perp}{2 (2 \pi)^3 \sqrt{x(1-x)}}\big[\Psi_{\pi(K)}(x,\textbf{k}_\perp, \uparrow, \uparrow)\ket{x P^+, \textbf{k}_\perp, \uparrow, \uparrow} \nonumber\\
&&+\Psi_{\pi(K)}(x,\textbf{k}_\perp, \downarrow, \downarrow)\ket{x P^+, \textbf{k}_\perp, \downarrow, \downarrow}+\Psi_{\pi(K)}(x,\textbf{k}_\perp, \downarrow, \uparrow)\nonumber\\
&&\ket{x P^+, \textbf{k}_\perp, \downarrow, \uparrow}+\Psi_{\pi(K)}(x,\textbf{k}_\perp, \uparrow, \downarrow)\ket{x P^+, \textbf{k}_\perp, \uparrow, \downarrow}\big].
\label{overlap}
\end{eqnarray}

\subsection{Form Factors}\label{emff}
The vector, tensor, and scalar quark FFs of pseudoscalar mesons can be calculated from the leading twist and sub-leading twist quark GPDs by integrating over longitudinal momentum fraction at zero skewness as \cite{QCDSF:2007ifr} 
\begin{eqnarray}
    F^q_v (Q^2)&=& \int dx F_1(x, \xi=0, -\Delta_\perp^2), \\
     F^q_T (Q^2)&=& \int dx H_1(x, \xi=0, -\Delta_\perp^2), \\
      F^q_S (Q^2)&=& \int dx E_2(x, \xi=0, -\Delta_\perp^2).
\end{eqnarray}
Here, $v$, $T$, and $S$ are noted for vector, tensor, and scalar, respectively. $F_1$, $H_1$, and $E_2$ correspond to 
vector, tensor, and scalar pseudoscalar meson quark GPDs, respectively with $F_1$ and $H_1$ being the twist-2 quark GPDs, wheras $E_2$ is the twist-3 quark GPD. $Q^2$ is the squared four-momentum transfer between the incoming and outgoing pion and kaon. $\Delta$ is the transverse momentum transfer between the initial and final state meson. The meson FFs can be calculated by taking into account each flavor of the meson and we have 
\begin{eqnarray}
    F^{\pi(K)}_{v,T,S}= e_q F^q_{v,T, S} + e_{\bar q} F^{\bar q}_{v,T, S},\label{ffq}
\end{eqnarray}
where $e_{q (\bar q)}$ is the charge of the quark (antiquark). The antiquark FFs can be calculated from antiquark GPDs, which are discussed further in the manuscript.
The matrix elements of the quark operators at a light-like separation are defined as GPDs \cite{Diehl:2003ny}. The valence quark GPDs for pseudoscalar mesons can be expressed through quark-quark correlators as \cite{Meissner:2008ay,Puhan:2024jaw}
\begin{equation}
      \Phi_q^{[\Gamma]}(x, \xi,-\Delta_\perp^2)=\frac{1}{2}\int\frac{\mathrm{d}z^-}{2\pi}e^{i \textbf{k}\cdot z} \langle \pi^+ (K^+) (P^{*}, S_z=0)|\bar\psi(-z/2)\Gamma\mathcal{W}\psi(z/2)|\pi^+ (K^+) (P,S_z=0) \rangle|_{z^+=z_\perp=0} ,\label{core}
\end{equation}
where $P^*$ and $P$ are the four vector momenta of the final and initial meson. $z=(z^+,z^-,z_\perp)$ and $\psi$ are the position four vector and the quark field operator. $\mathcal{W}$ is the Wilson line which connects the quark field operator from $-z/2$ to $z/2$ and  ensures the color gauge invariance. The variables $\Delta= (\Delta^+,\Delta^-,\Delta_\perp)=(P^*-P)$ and $\xi=-\Delta^+/(P^*+P)$ are the momentum transfer between the initial and final state meson and the skewness parameter. Here, $Q^2=\Delta^2=-\Delta_\perp^2$ as $\Delta^+=0$ for our case. In this work, we mainly focus on the transverse momentum transferred $\Delta_\perp^2$ which gives rise to FFs at $\xi=0$. $\Gamma$ is the gamma matrix specific for each GPDs. $\Gamma=\gamma^+, \iota\sigma^{j+}\gamma_5$, and $1$ corresponds to vector, tensor, and scalar quark GPDs, respectively. These GPDs can be expressed in the form of $\Phi_q^{[\Gamma]}(x, \xi,-\Delta_\perp^2)$ as 
\begin{eqnarray}
    F_1(x, \xi=0, -\Delta_\perp^2)&=& \Phi_q^{[\gamma^+]}(x, \xi,-\Delta_\perp^2), \\
     H_1(x, \xi=0, -\Delta_\perp^2) &=& -\frac{\iota \mathcal{M}_{\pi(K)}}{\epsilon^{ij} \Delta_\perp^i}\Phi_q^{[\iota \sigma^{j+}\gamma_5]}(x, \xi,-\Delta_\perp^2), \\
     E_2(x, \xi=0, -\Delta_\perp^2) &=& \frac{P^+}{\mathcal{M}_{\pi(K)}}\Phi_q^{[1]}(x, \xi,-\Delta_\perp^2).
\end{eqnarray}
Now using the pion and kaon meson Fock-state of Eq. (\ref{overlap}) in the quark-quark correlator of Eq. (\ref{core}), the overlap form of different FFs with all possibilities of helicities are found to be 
\begin{eqnarray}
    F_v^q(Q^2)&=& \int \frac{dx d^2\bfk}{16 \pi^3} \Bigg[\Psi^*_{\pi(K)}(x,\bfk^{\prime\prime},\uparrow,\uparrow) \Psi_{\pi(K)}(x,\bfk^{\prime},\uparrow,\uparrow)+\Psi^*_{\pi(K)}(x,\bfk^{\prime\prime},\uparrow,\downarrow) \Psi_{\pi(K)}(x,\bfk^{\prime},\uparrow,\downarrow) \nonumber\\
    &&+\Psi^*_{\pi(K)}(x,\bfk^{\prime\prime},\downarrow,\uparrow) \Psi_{\pi(K)}(x,\bfk^{\prime},\downarrow,\uparrow)+\Psi^*_{\pi(K)}(x,\bfk^{\prime\prime},\downarrow,\downarrow) \Psi_{\pi(K)}(x,\bfk^{\prime},\downarrow,\downarrow)\Big], \\
     F_T^q(Q^2)&=& \int \frac{dx d^2\bfk}{16 \pi^3} \mathcal{M}_{\pi(K)}\Bigg[\iota \Big(\Psi^*_{\pi(K)}(x,\bfk^{\prime\prime},\downarrow,\uparrow) \Psi_{\pi(K)}(x,\bfk^{\prime},\uparrow,\uparrow)+\Psi^*_{\pi(K)}(x,\bfk^{\prime\prime},\downarrow,\downarrow) \Psi_{\pi(K)}(x,\bfk^{\prime},\uparrow,\downarrow) \Big)\nonumber\\
    &&-\iota \Big(\Psi^*_{\pi(K)}(x,\bfk^{\prime\prime},\uparrow,\uparrow) \Psi_{\pi(K)}(x,\bfk^{\prime},\downarrow,\uparrow)+\Psi^*_{\pi(K)}(x,\bfk^{\prime\prime},\uparrow,\downarrow) \Psi_{\pi(K)}(x,\bfk^{\prime},\downarrow,\downarrow)\Big)\Big], \\
    F_S^q(Q^2)&=& \int \frac{dx d^2\bfk}{\mathcal{M}_{\pi(K)}16 \pi^3} \Bigg[\frac{m_q}{x}\Big(\Psi^*_{\pi(K)}(x,\bfk^{\prime\prime},\uparrow,\uparrow) \Psi_{\pi(K)}(x,\bfk^{\prime},\uparrow,\uparrow)+\Psi^*_{\pi(K)}(x,\bfk^{\prime\prime},\uparrow,\downarrow) \Psi_{\pi(K)}(x,\bfk^{\prime},\uparrow,\downarrow) \nonumber\\
    &&+\Psi^*_{\pi(K)}(x,\bfk^{\prime\prime},\downarrow,\uparrow) \Psi_{\pi(K)}(x,\bfk^{\prime},\downarrow,\uparrow)+\Psi^*_{\pi(K)}(x,\bfk^{\prime\prime},\downarrow,\downarrow) \Psi_{\pi(K)}(x,\bfk^{\prime},\downarrow,\downarrow)\Big)\nonumber\\
    &&+\frac{(\Delta_1+\iota \Delta_2)}{2 x}\Big(\Psi^*_{\pi(K)}(x,\bfk^{\prime\prime},\downarrow,\uparrow) \Psi_{\pi(K)}(x,\bfk^{\prime},\uparrow,\uparrow)+\Psi^*_{\pi(K)}(x,\bfk^{\prime\prime},\downarrow,\downarrow) \Psi_{\pi(K)}(x,\bfk^{\prime},\uparrow,\downarrow)\Big)\nonumber\\
    &&-\frac{(\Delta_1-\iota\Delta_2)}{2 x}\Big(\Psi^*_{\pi(K)}(x,\bfk^{\prime\prime},\uparrow,\uparrow) \Psi_{\pi(K)}(x,\bfk^{\prime},\downarrow,\uparrow)+\Psi^*_{\pi(K)}(x,\bfk^{\prime\prime},\uparrow,\downarrow) \Psi_{\pi(K)}(x,\bfk^{\prime},\downarrow,\downarrow)\Big)\Big].
\end{eqnarray}
Here, $\bfk^{\prime \prime}=\bfk-\Delta_\perp/2$ and $\bfk^{\prime}=\bfk+\Delta_\perp/2$ are the transverse momenta carried by the quark in the final and initial state meson. For this work, we have chosen the symmetric frame or Breit frame for our calculations. 
For tensor FF, we have taken $j=2$. Taking $j=1$ does not affect the final results. The explicit forms of quark FFs for pion by using the spin and momentum space wave function are expressed as 
\begin{eqnarray}
    F_v^{q(\pi)}(Q^2)&=&\int \frac{d x d^2\bfk}{16 \pi^3} \Big[\bfk^2-(1-x)^2\frac{\Delta^2_\perp}{4}+(m_u(1-x)+x m_{\bar d})^2\Big]\frac{\psi^*_{\pi}(x,\bfk^{\prime\prime})\psi_\pi(x,\bfk^\prime)}{\omega_{\pi}^{\prime \prime}\omega_\pi^\prime}, \\
   F_T^{q(\pi)}(Q^2)&=&\int \frac{d x d^2\bfk}{16 \pi^3} \Big[\mathcal{M}_{\pi}(1-x)(m_u(1-x)+m_{\bar d}x)]\frac{\psi^*_{\pi}(x,\bfk^{\prime\prime})\psi_\pi(x,\bfk^\prime)}{\omega_\pi^{\prime \prime}\omega_\pi^\prime}, \end{eqnarray}
\begin{equation}
    F_S^{q(\pi)}(Q^2)=\int \frac{d x d^2\bfk}{16 \pi^3}\frac{2 m_u}{x \mathcal{M}_\pi} \Big[\bfk^2-(1-x)^2\frac{\Delta_\perp^2}{4}+(m_u(1-x)+x m_{\bar d})^2+m_u (1-x) \Delta_\perp^2\Bigg]\frac{\psi^*_{\pi}(x,\bfk^{\prime\prime})\psi_\pi(x,\bfk^\prime)}{\omega_\pi^{\prime \prime}\omega_\pi^\prime},
\end{equation}
where 
\begin{eqnarray}
   \omega_\pi^{\prime \prime}&=& (\mathcal{M}^{\prime \prime}_\pi+m_u+m_{\bar d})\sqrt{x(1-x)[\mathcal{M}^{\prime \prime2}_{\pi}-(m_u-m_{\bar d})^2]}, \\
   \omega_\pi^{\prime}&=& (\mathcal{M}^{\prime}_\pi+m_u+m_{\bar d})\sqrt{x(1-x)[\mathcal{M}^{\prime2}_{\pi}-(m_u-m_{\bar d})^2]}.
\end{eqnarray}
With bound state pion mass
\begin{eqnarray*}
    \mathcal{M}_\pi^{\prime \prime }&=&\sqrt{\frac{(\bfk-\frac{\Delta_\perp}{2})^2+m^2_u}{x}+\frac{(\bfk-\frac{\Delta_\perp}{2})^2+m^2_{\bar d}}{1-x}}, \nonumber \\  \mathcal{M}_\pi^{\prime }&=&\sqrt{\frac{(\bfk+\frac{\Delta_\perp}{2})^2+m^2_u}{x}+\frac{(\bfk+\frac{\Delta_\perp}{2})^2+m^2_{\bar d}}{1-x}}.
\end{eqnarray*}
Similarly, the explicit form of the quark FFs of the kaon are found to be 
\begin{eqnarray}
F_v^{q(K)}(Q^2)&=&\int \frac{d x d^2\bfk}{16 \pi^3} \Big[\bfk^2-(1-x)^2\frac{\Delta^2_\perp}{4}+(m_u(1-x)+x m_{\bar s})^2\Big]\frac{\psi^*_K(x,\bfk^{\prime\prime})\psi_K(x,\bfk^\prime)}{\omega_{K}^{\prime \prime}\omega_K^\prime}, \\
   F_T^{q(K)}(Q^2)&=&\int \frac{d x d^2\bfk}{16 \pi^3} \Big[\mathcal{M}_{K}(1-x)(m_u(1-x)+m_{\bar s}x)]\frac{\psi^*_{K}(x,\bfk^{\prime\prime})\psi_K(x,\bfk^\prime)}{\omega_K^{\prime \prime}\omega_K^\prime}, \\
    F_S^
    {q(K)}(Q^2)&=&\int \frac{d x d^2\bfk}{16 \pi^3}\frac{1}{x \mathcal{M}_K} \Big[2 m_u\Bigg(\bfk^2-(1-x)^2\frac{\Delta_\perp^2}{4}+(m_u(1-x)+x m_{\bar s})^2\Bigg)\nonumber\\
    &&+2 ((1-x)m_u+x m_{\bar s})^2(1-x) \Delta_\perp^2\Bigg]\frac{\psi^*_{K}(x,\bfk^{\prime\prime})\psi_K(x,\bfk^\prime)}{\omega_K^{\prime \prime}\omega_K^\prime}.
\end{eqnarray}
The antiquark FFs can be calculated from antiquark GPDs of the meson. The antiquark GPDs can be calculated from the quark GPDs as
\begin{eqnarray}
   {\rm quark ~~ GPD}(x,\xi,-\Delta_\perp^2,m_q,m_{\bar q}) ={\rm antiquark ~~ GPD} (1-x,\xi,-\Delta_\perp^2,m_{\bar q},m_{q}).
\end{eqnarray}
Both the quark and antiquark FFs will give rise to meson FFs as already mentioned in Eq. (\ref{ffq}). 
\section{Results and Discussions}\label{result}
 \begin{figure}[ht]
		\centering
		\begin{minipage}[c]{1\textwidth}\begin{center}
				(a)\includegraphics[width=.45\textwidth]{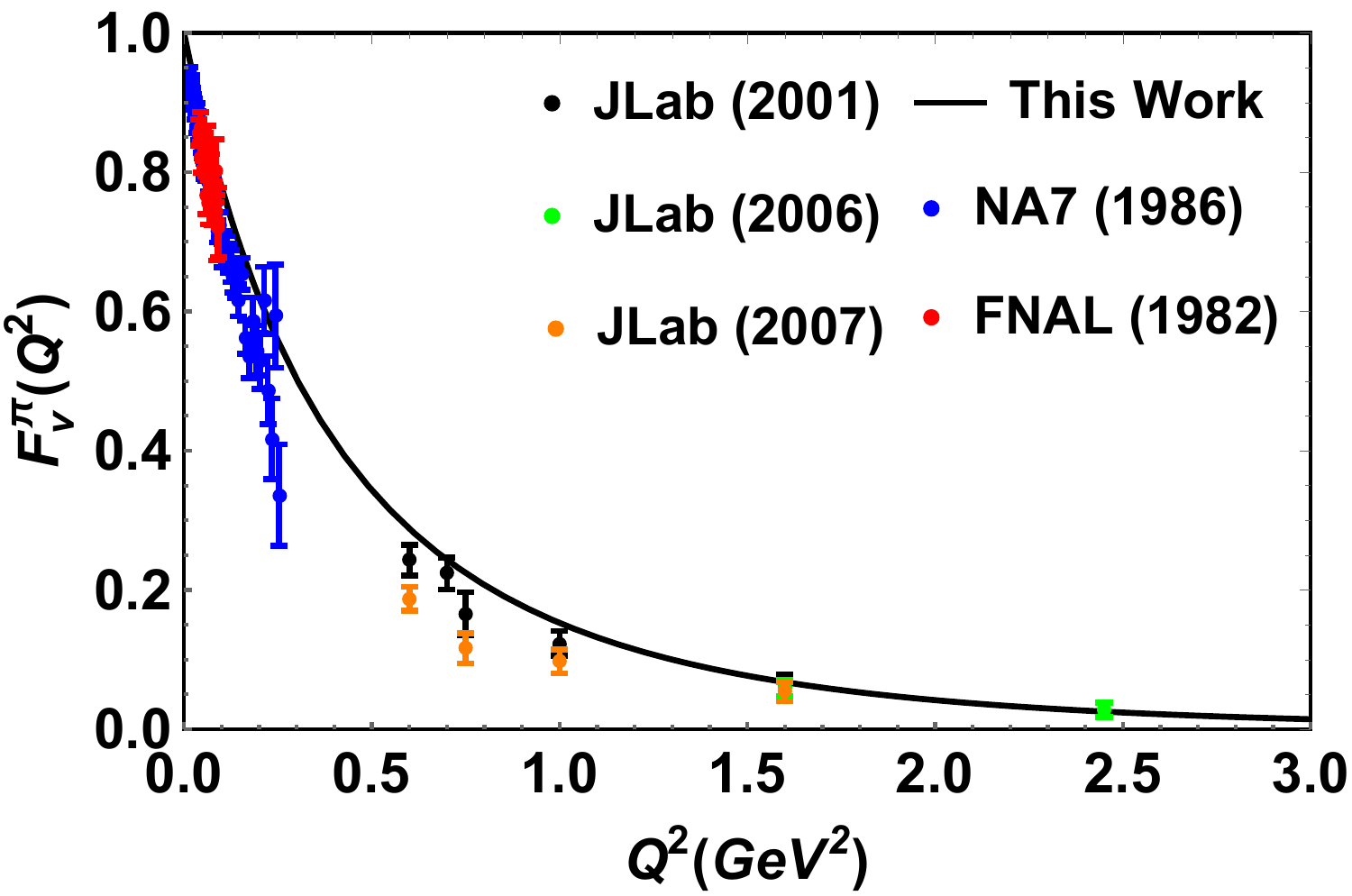}
				(b)\includegraphics[width=.45\textwidth]{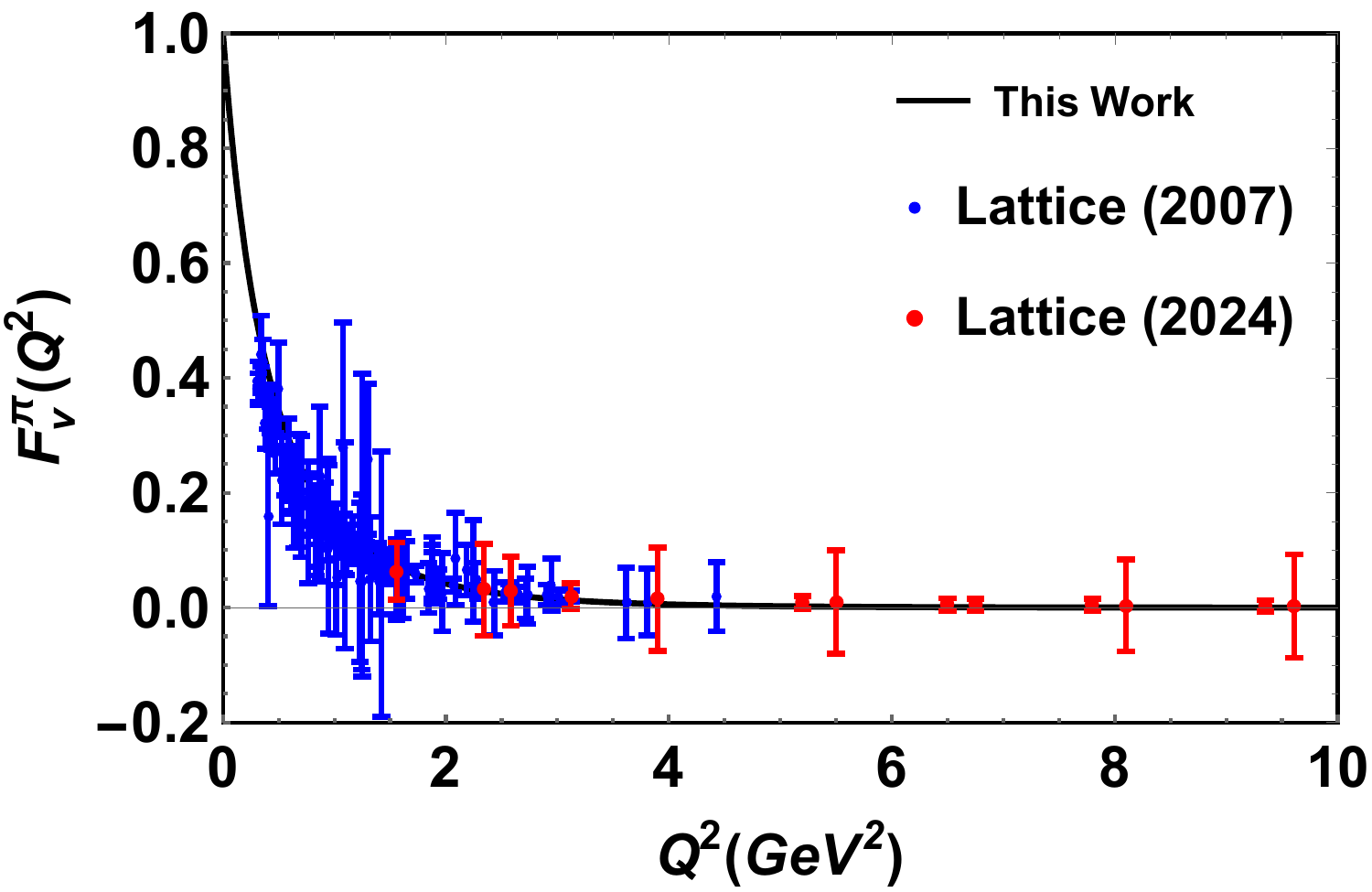}
			\end{center}
		\end{minipage}
		\begin{minipage}[c]{1\textwidth}\begin{center}
				(c)\includegraphics[width=.45\textwidth]{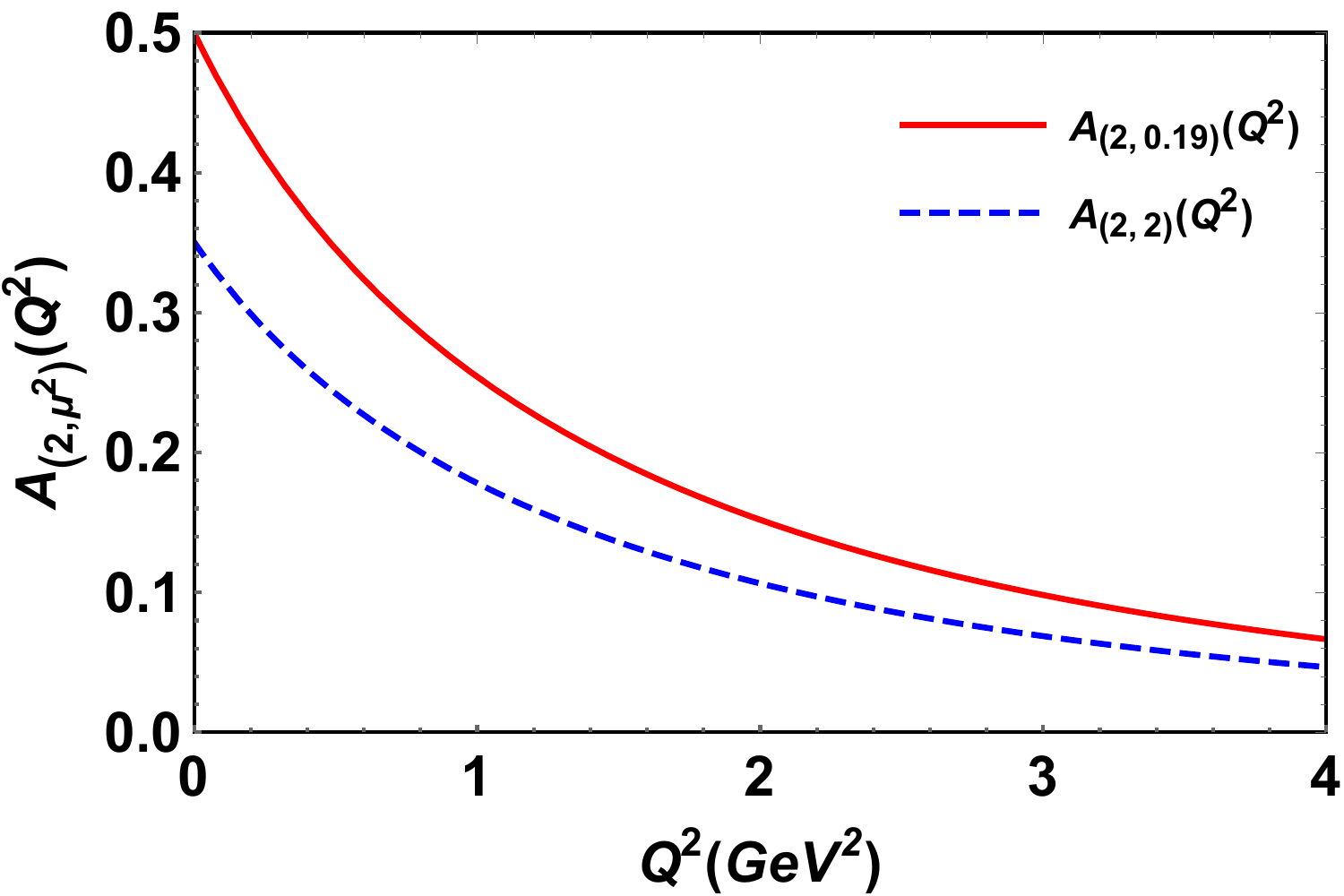}
                (d)\includegraphics[width=.45\textwidth]{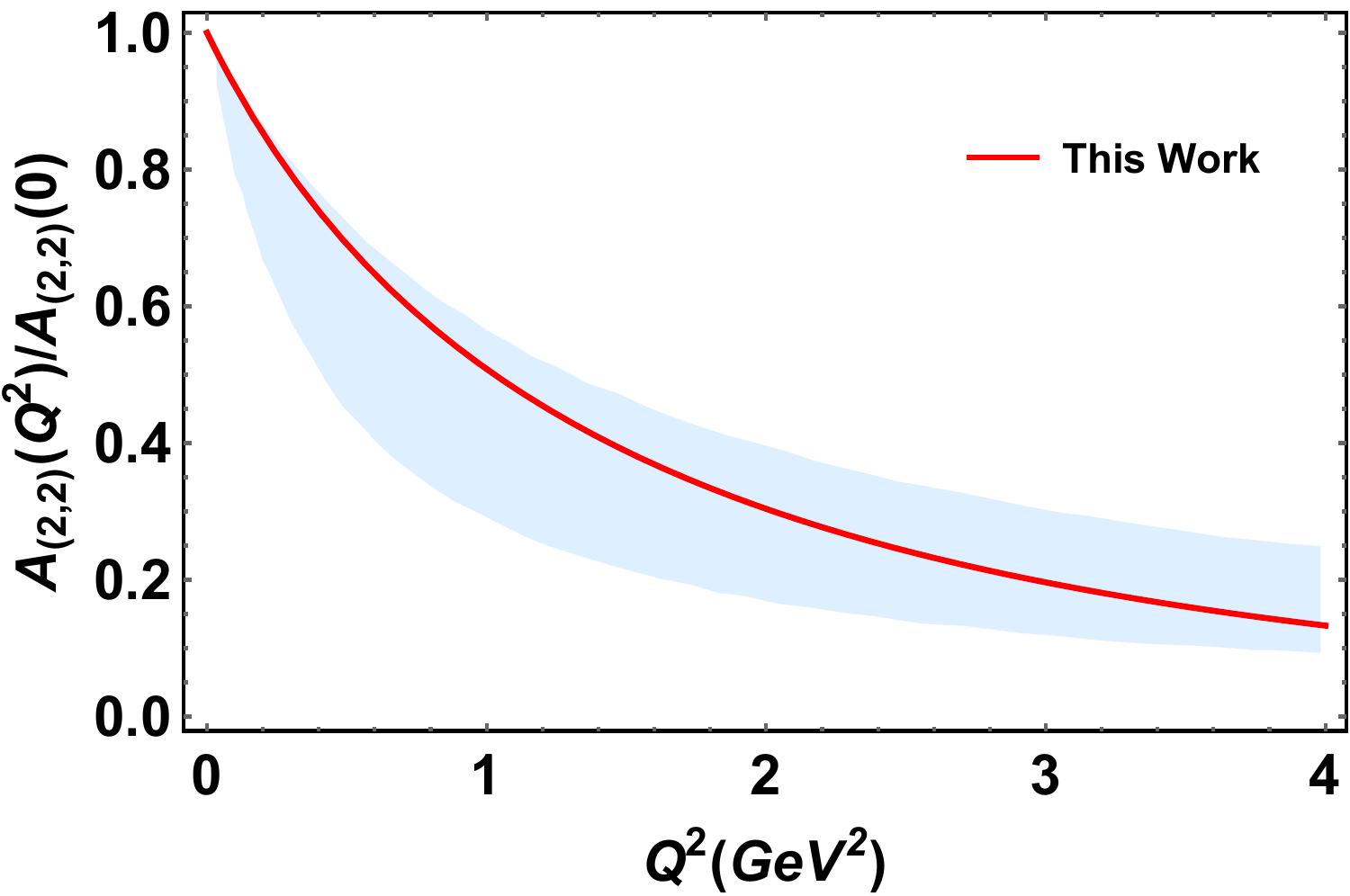}
                \end{center}
		\end{minipage}
		\caption{(Color online) 
        The vector $F_v^{\pi}(Q^2)$ FF has been plotted with respect to $Q^2$ for the case of pion. (a) The vector FF have been compared with NA7 \cite{NA7:1986vav}, FNAL \cite{Dally:1982zk}, JLab (2001) \cite{JeffersonLabFpi:2000nlc}, JLab (2006) \cite{JeffersonLabFpi-2:2006ysh}, and JLab (2007) \cite{JeffersonLabFpi:2007vir} upto $Q^2=3$ GeV$^2$. (b) Pion vector FF has been compared with lattice simulation results \cite{QCDSFUKQCD:2006gmg,Ding:2024lfj}. (c) The second moment of pion vector FF has been plotted at model scale $\mu^2=0.19$ GeV$^2$ and at $\mu^2=2$ GeV$^2$. (d) The normalized second moment pion vector FF has been compared with lattice simulation results (shaded region) \cite{Brommel:2007zz}.}
		\label{vector}
	\end{figure}
For the numerical predictions of FFs of pion and kaon, we have considered the LCQM. In this model, we need only two input parameters: quark masses ($m_q$) and harmonic scale parameter ($\beta_{\pi(K)}$) for the calculations. $m_{u(\bar d)}$ and $\beta_\pi$ are taken to be $0.2$ and $0.410$ respectively, for pion \cite{Kaur:2020vkq} whereas $m_{u(\bar s)}$ and $\beta_{K}$ are taken to be $0.2 (0.556)$ and $0.405$ for the case of kaon \cite{Kaur:2020vkq,Puhan:2024jaw}. These input parameters have been calculated by fitting the mass spectra and they successfully produce the decay constant, PDFs, and charge radii quite well compared to other parameter values \cite{Kaur:2020vkq,Puhan:2024jaw}. The calculated FFs of pion and kaon are functions of only $Q^2=-\Delta_\perp^2$ (GeV$^2$). The EMFF or vector FF of pion and kaon have already been studied in our previous work \cite{Puhan:2024jaw}, however, for the sake of completeness, we have plotted and discussed them here also. We have plotted the vector FF $F_v^{\pi}(Q^2)$ of the pion in Fig. \ref{vector} up to different $Q^2$ (GeV$^2$). $F_v^{\pi}(Q^2)$ is found to be a smoothly decreasing function of $Q^2$ and almost saturates after $Q^2=4$ GeV$^2$. In Figs. \ref{vector} (a) and (b), we have compared our vector FF $F_v^{\pi}(Q^2)$ of pion results with available experimental results \cite{NA7:1986vav,Dally:1982zk,JeffersonLabFpi:2000nlc,JeffersonLabFpi-2:2006ysh,JeffersonLabFpi:2007vir} up to $Q^2=3$ GeV$^2$ and lattice simulation data \cite{QCDSFUKQCD:2006gmg,Ding:2024lfj} up to $Q^2=10$ GeV$^2$ respectively. The vector FF of the pion is found to be slightly deviated from experimental results at low $Q^2$, and at high $Q^2$, the result is found to match with JLab data. However, the vector FF result is found to exactly match the latest lattice simulation data at high $Q^2$\cite{Ding:2024lfj}. The vector FFs of pion and kaon also obey the sum rule $F_v(Q^2=0)=1$. In Fig. \ref{vector} (c), we have plotted the second moment of vector FF $A_2(Q^2)$ at model scale  $\mu^2=0.19$ as well as at scale $\mu^2=2$ for the case of pion.
\par
\begin{figure}[ht]
		\centering
		\begin{minipage}[c]{1\textwidth}\begin{center}
				(a)\includegraphics[width=.45\textwidth]{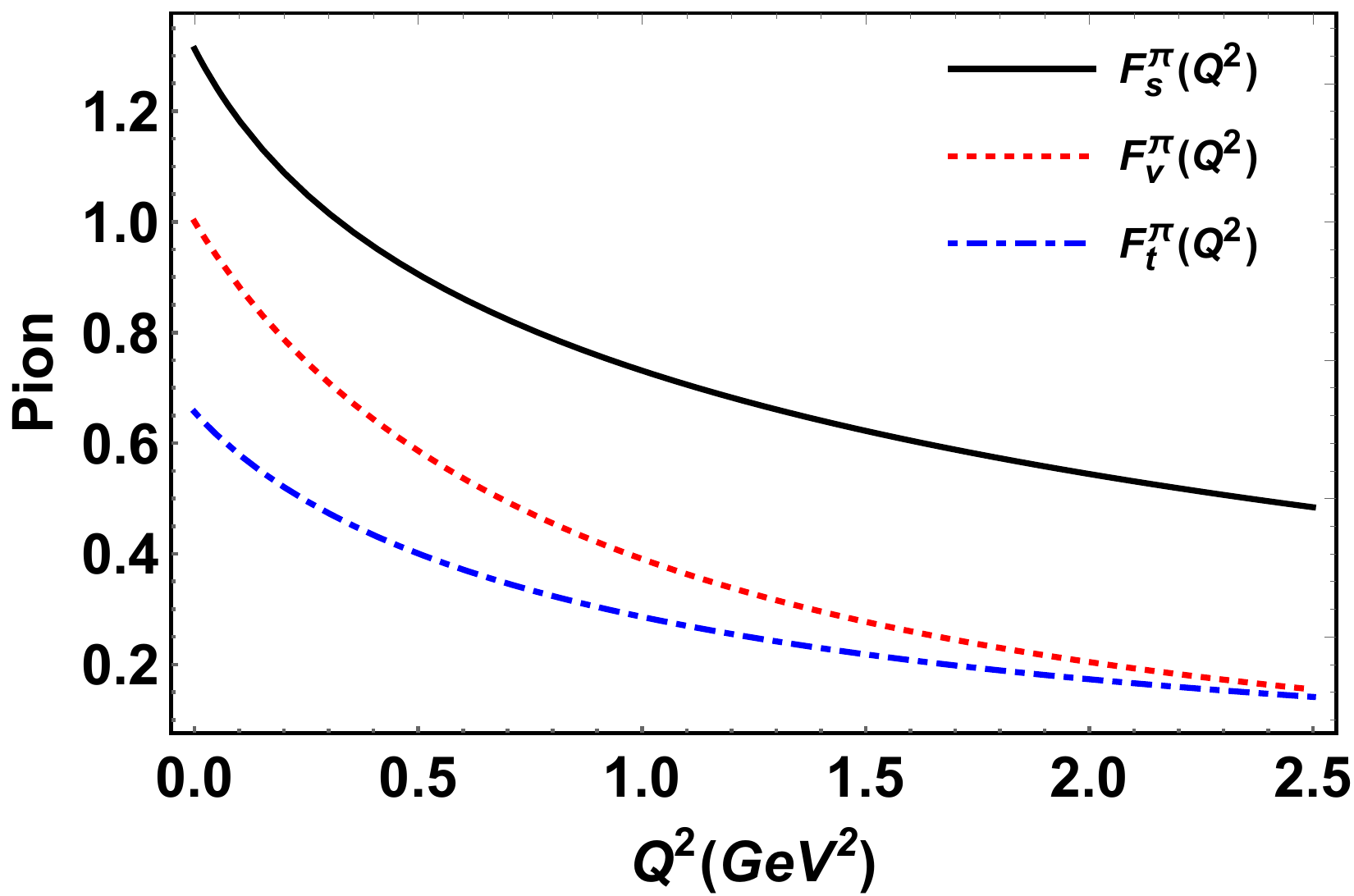}
				(b)\includegraphics[width=.45\textwidth]{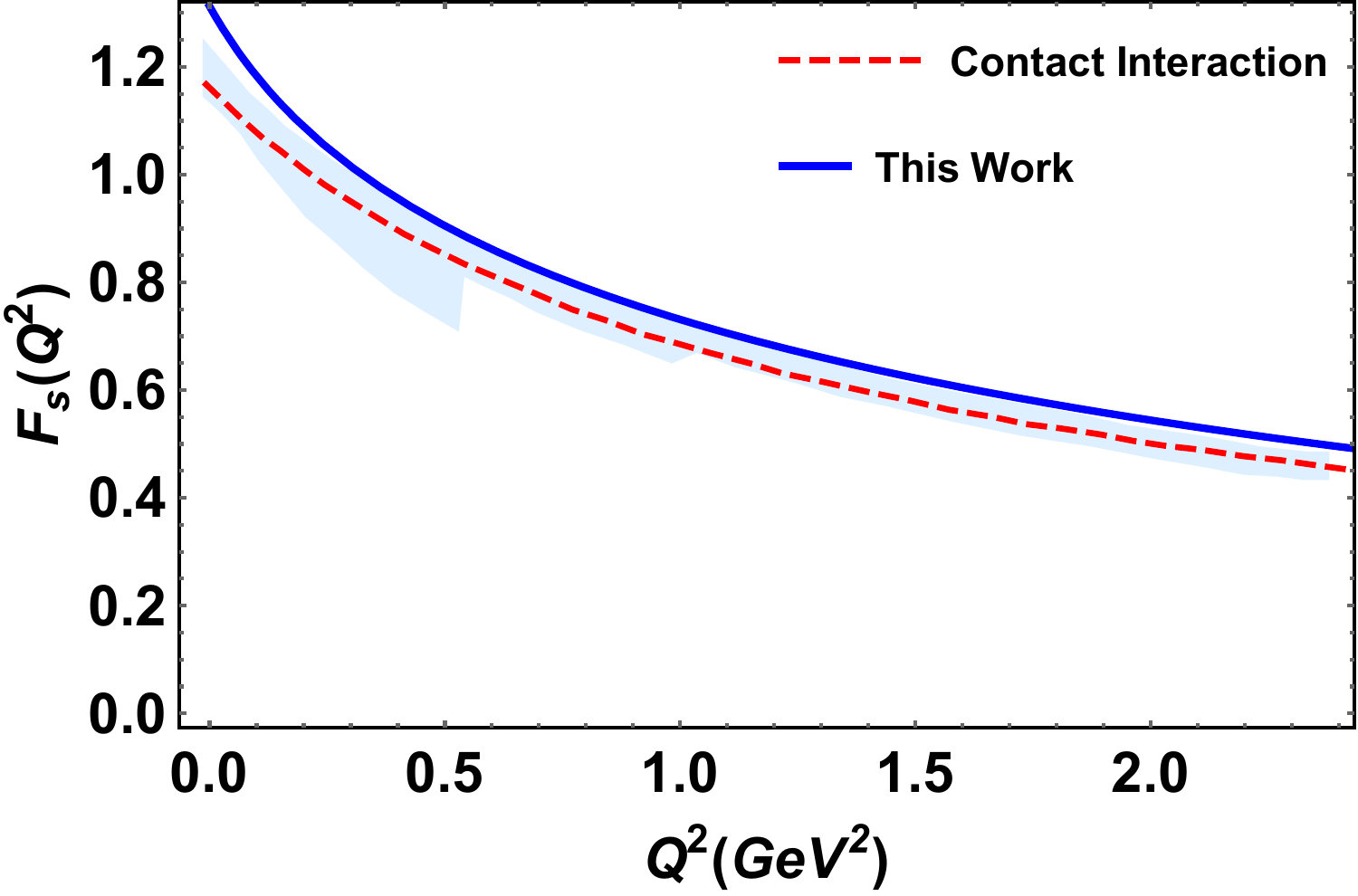}
			\end{center}
		\end{minipage}
		\begin{minipage}[c]{1\textwidth}\begin{center}
				(c)\includegraphics[width=.45\textwidth]{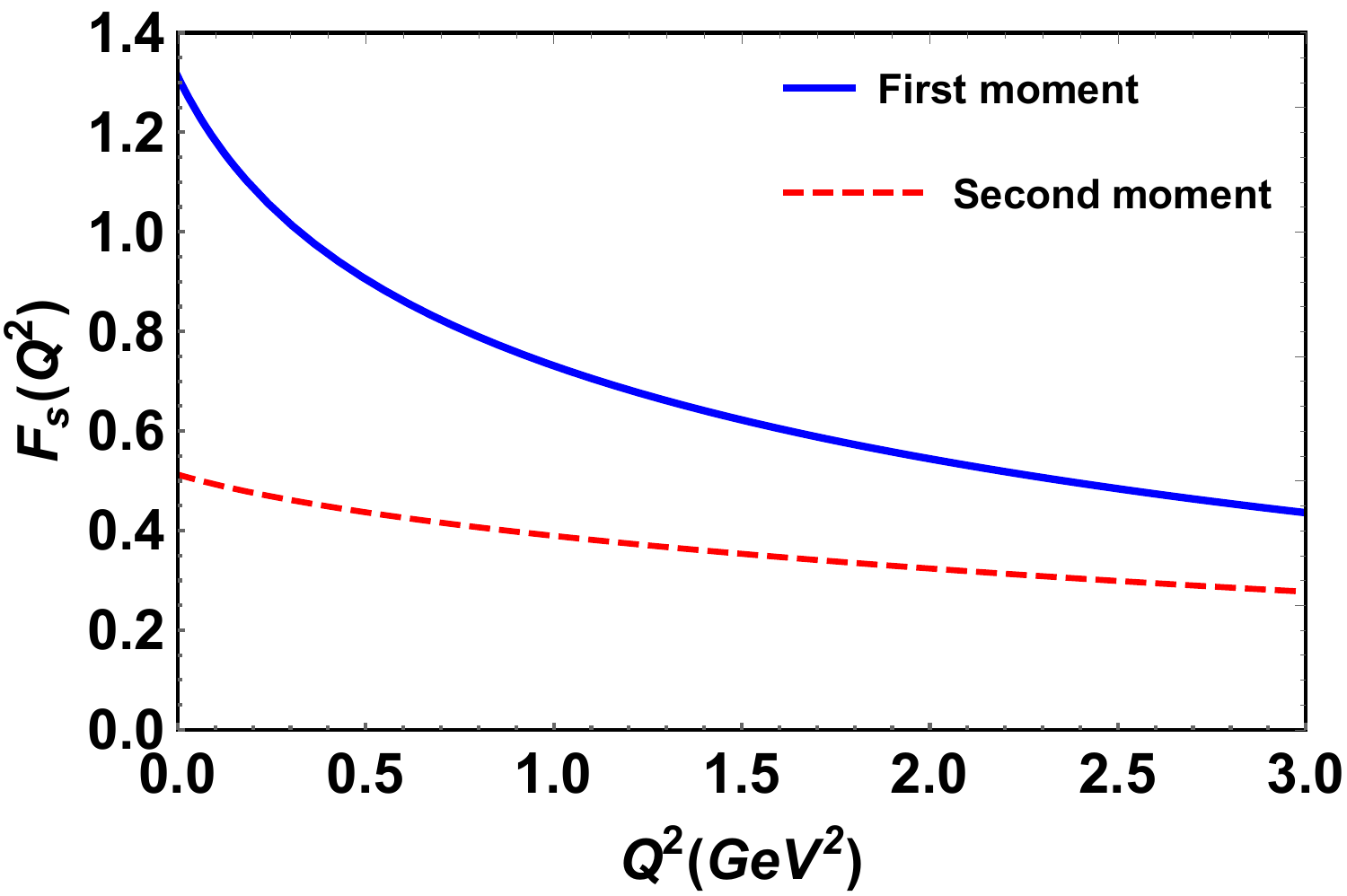}
                \end{center}
		\end{minipage}
		\caption{(Color online) 
      (a) The scalar ($F_S^{\pi}(Q^2)$), vector ($F_v^{\pi}(Q^2)$), and tensor ($F_T^{\pi}(Q^2)$) FF of the pion have been plotted with respect to $Q^2$ GeV$^2$. (b) The scalar FF ($F_S(Q^2)$) of our work has been compared with only available lattice simulation data (shaded region) \cite{Alexandrou:2021ztx} and theoretical model (contact interaction) predictions \cite{Wang:2022mrh}. (c) The first and second moments of the scalar FF of the pion have been plotted with respect to $Q^2$ GeV$^2$.}
		\label{scalarff}
	\end{figure}

 The second moment of the pion and kaon is calculated from second moment of GPDs as
\begin{eqnarray}
    A_2(Q^2)=\int dx x F_1(x,0,-\Delta_\perp^2), \ \ \ \ B_2(Q^2)=\int dx x H_1(x,0,-\Delta_\perp^2).
\end{eqnarray}
Here, $A$ and $B$ denote the second moment of the vector and tensor FFs, respectively. The evolution of FFs has been done by using the next-to-leading (NLO) order kernel as follows \cite{Son:2015bwa,Fanelli:2016aqc,Belitsky:2005qn}

\begin{equation}
  \label{eq:RG}
FF(Q^2,\mu^2) =
\left(\frac{\alpha(\mu^2)}{\alpha({\mu_0^2})}\right)^{4/27}
\left[ 1-\frac{337}{486\pi}(\alpha(\mu_0^2)-\alpha(\mu^2))
\right]FF(Q^2,\mu_0^2),
\end{equation}
with the NLO strong coupling constant
\begin{equation}
\alpha(\mu^2)=
\frac{4\pi}{9\ln(\mu^2/\Lambda_{\mathrm{QCD}}^2)}
\left[
1-\frac{64}{81}\frac{\ln\ln(\mu^2/\Lambda_{\mathrm{QCD}}^2)}{
  \ln(\mu^2/\Lambda_{\mathrm{QCD}}^2)}
\right].
\end{equation}
    \begin{figure}[ht]
		\centering
		\begin{minipage}[c]{1\textwidth}\begin{center}
				(a)\includegraphics[width=.45\textwidth]{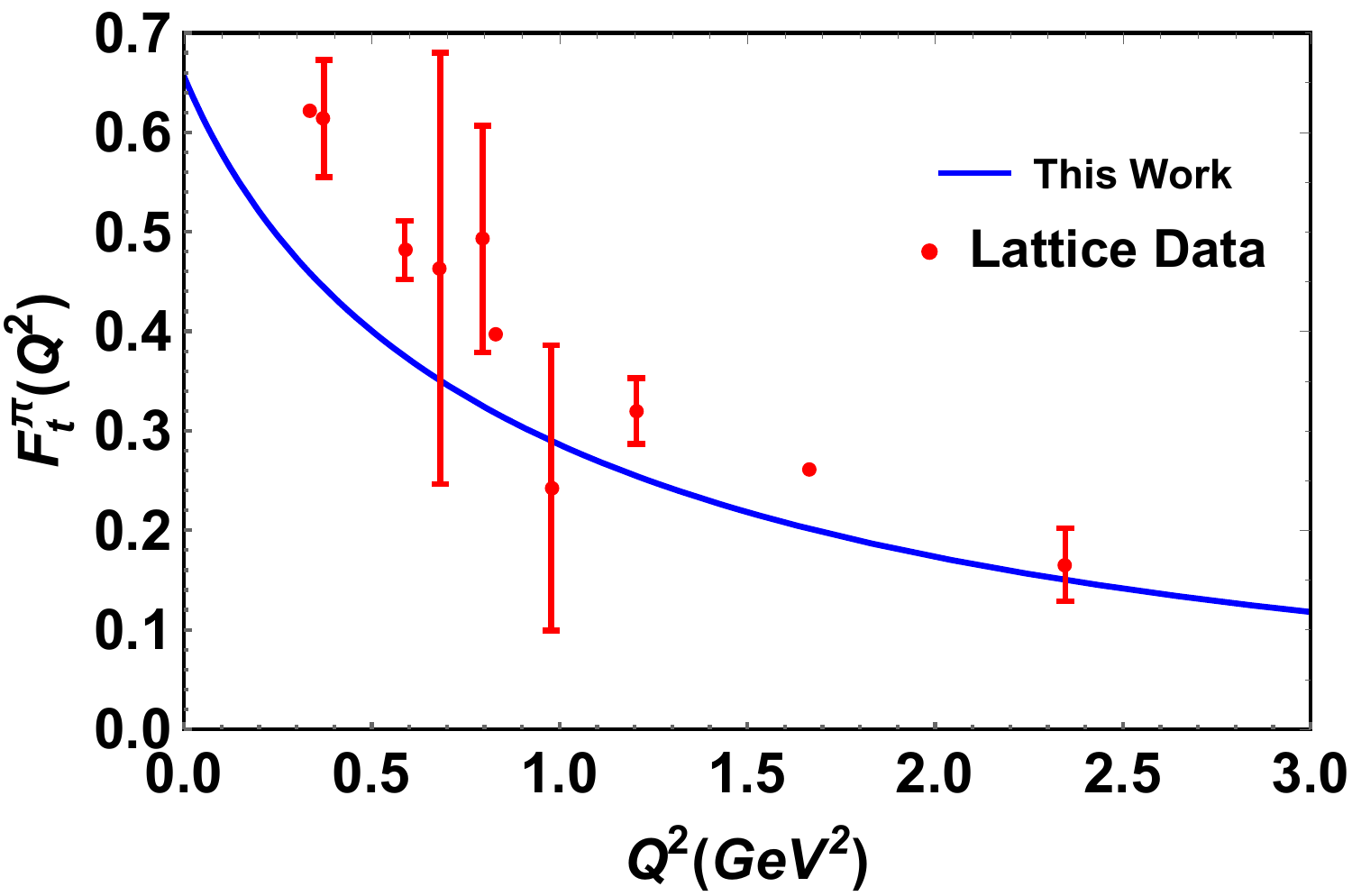}
				(b)\includegraphics[width=.45\textwidth]{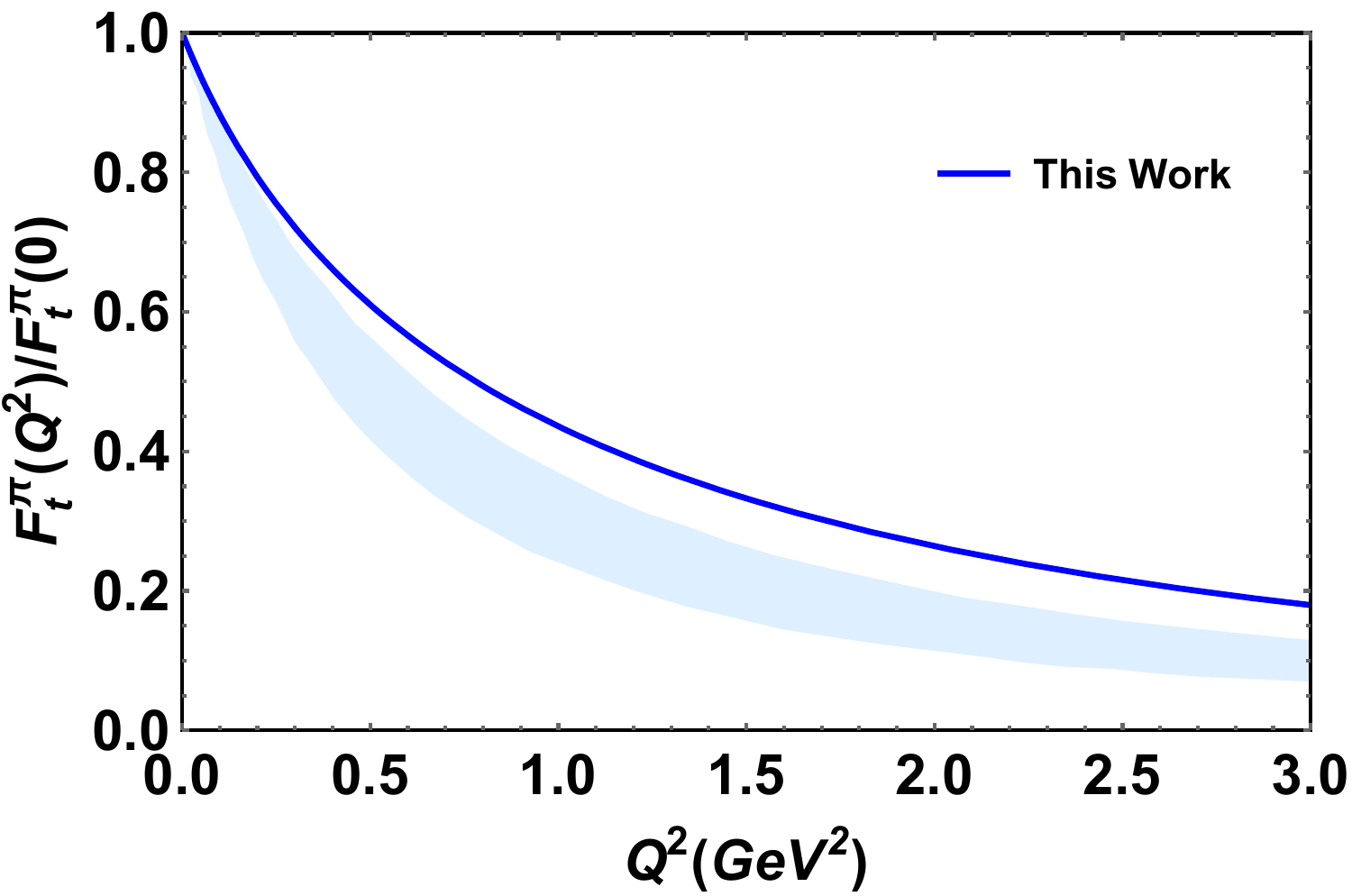}
			\end{center}
		\end{minipage}
		\begin{minipage}[c]{1\textwidth}\begin{center}
				(c)\includegraphics[width=.45\textwidth]{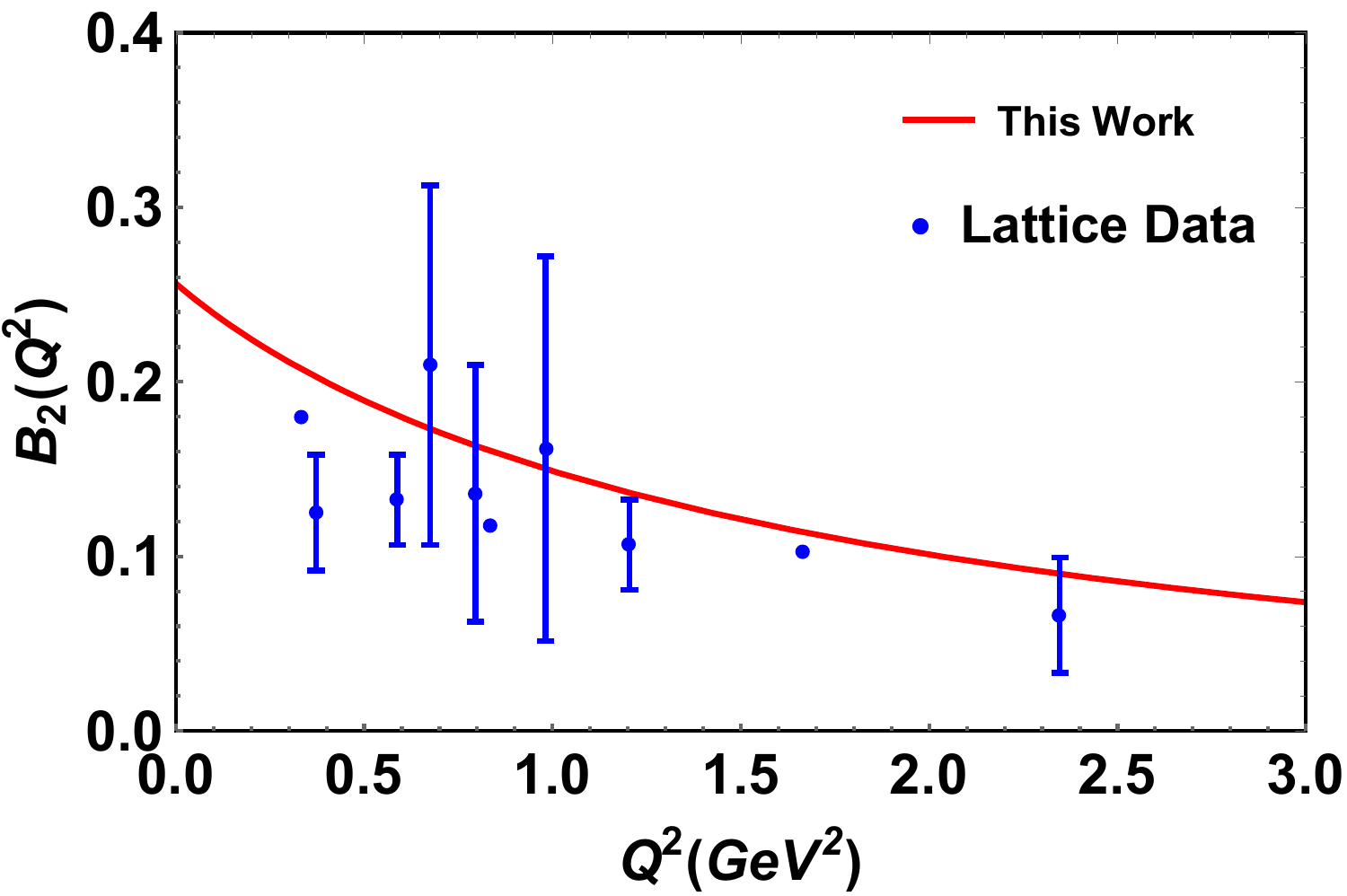}
                (d)\includegraphics[width=.45\textwidth]{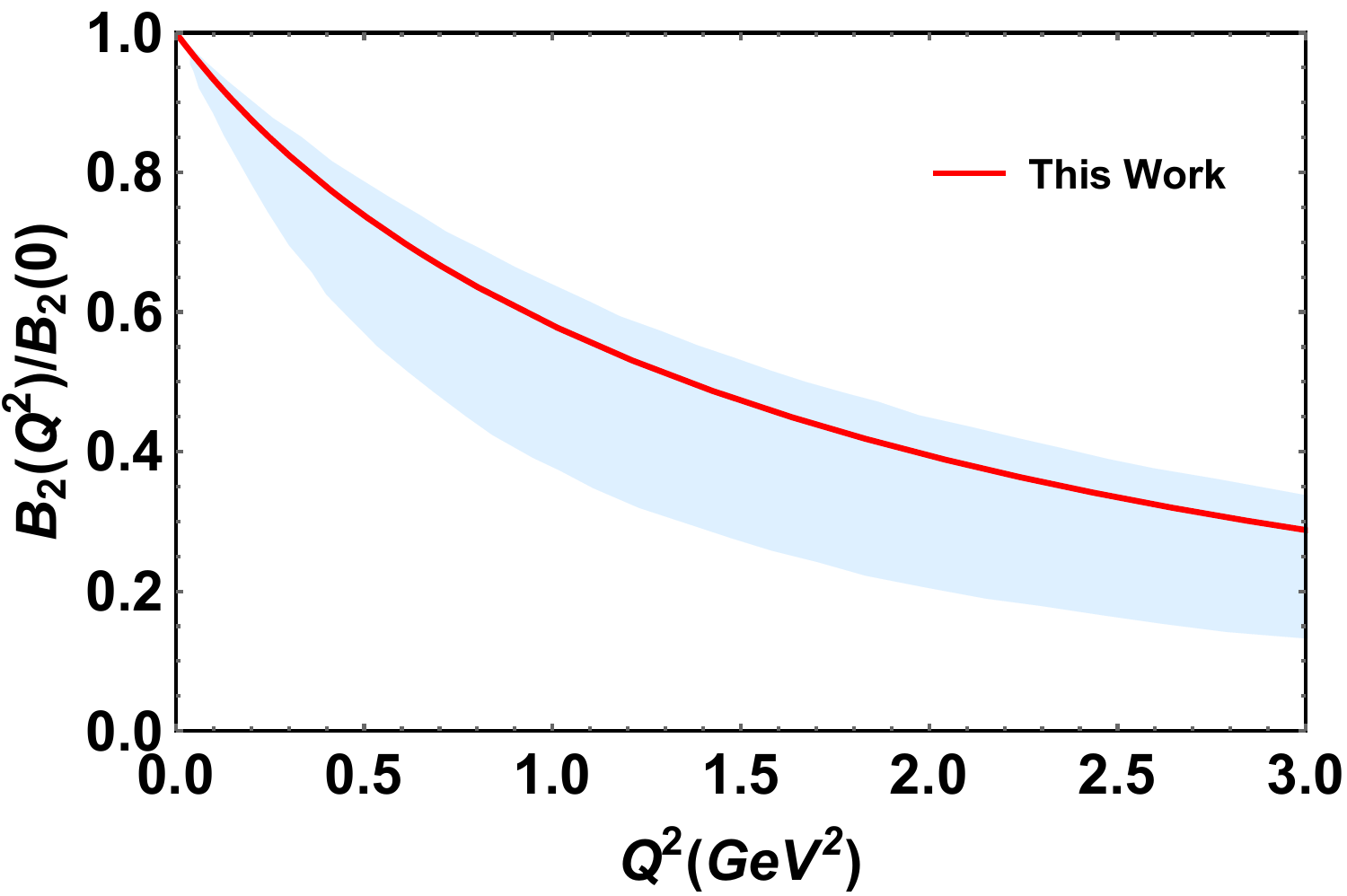}
                \end{center}
		\end{minipage}
		\caption{(Color online) The tensor FF of pion has been plotted with respect to $Q^2$ GeV$^2$. The first ($F_t^{\pi}(Q^2)$) and second moment ($B_2(Q^2)$) of the tensor FF of the pion have been compared with lattice simulation data \cite{QCDSF:2007ifr} in (a) and (c), respectively. The normalized first and second moment of the tensor FF of the pion have been compared with lattice simulation data (shaded region) shown in Refs. \cite{Fanelli:2016aqc,Brommel:2007zz} in (b) and (d), respectively.}
		\label{tensor}
	\end{figure}
The model scale $\mu_0^2$ of LCQM has been taken as $0.19$ GeV$^2$ \cite{Kaur:2020emh} for this work. $\Lambda_{\mathrm{QCD}}^2$ is the QCD scale, which is taken to be $0.30$ GeV$^2$. In Fig. \ref{vector} (c), we observed that the evolved $A_2(Q^2)$ is reduced by $30$\% at $Q^2=0$ compared to the un-evolved FF at the model scale. The normalized second moment FF $A_{(2,2)}$ at $\mu^2=2$ GeV$^2$ has been compared with only available lattice simulation data \cite{Brommel:2007zz} and found to be in sync with it. The tensor FF is the result of polarized GPDs and spin-flip wave function. As there is no experimental result available for the tensor FF of the pion, so we have tried to compare our results with available lattice simulation results and theoretical predictions. The first moment $F_T(Q^2)$ and second moment $B_2(Q^2)$ of tensor FFs have been presented and compared with lattice simulation data \cite{QCDSF:2007ifr} in Figs. \ref{tensor} (a) and (c), respectively. Our $F_T(Q^2)$ FF is found to have a similar kind of behavior to contact interaction results \cite{Wang:2022mrh}, instanton vacuum \cite{Nam:2010pt}, chiral quark model \cite{Broniowski:2010nt}, and lattice simulation results \cite{Alexandrou:2021ztx}, but has a higher magnitude in distribution compared to all at low $Q^2$. At $Q^2=0$, our $F_T^\pi(Q^2)$ is coming out to be $0.655$ compared to $0.856$ in \cite{QCDSF:2007ifr}, $0.1612$ in \cite{Fanelli:2016aqc}, $0.362$ in \cite{Wang:2022mrh}, and $0.4$ in \cite{Alexandrou:2021ztx}, respectively. While the evolved $F_T^\pi(Q^2=0)$ FF at $\mu^2=2$ GeV$^2$ is found to be $0.460$ of our model compared to $0.216$ in Ref. \cite{Nam:2010pt}, $0.149$ in Ref. \cite{Broniowski:2010nt}, and $0.216$ in Ref. \cite{Brommel:2007zz}. In Figs. \ref{tensor} (b) and (d), we have compared our normalized first and second moment tensor FFs with only available lattice simulation data \cite{Brommel:2007zz}. The normalized second tensor FF $B_2(Q^2)$ perfectly matches with the lattice simulation result, while the normalized $F_T^{\pi} (Q^2)$ has a higher distribution at $Q^2\ge 0.3$ compared to the lattice simulation data \cite{Brommel:2007zz}. The scalar FF $F_S(Q^2)$ has been derived from twist-$3$ scalar GPDs for the case of pion and kaon. This FF has been plotted with respect to $Q^2$ for the case of pion in Fig. \ref{scalarff}. Similar to tensor FF, there is no experimental data available for the case of scalar FFs and very less studies have been reported for this FF \cite{Wang:2022mrh,Alexandrou:2021ztx}. The scalar FF of the pion is found to have a higher distribution compared to vector and tensor FFs, as shown in Fig. \ref{scalarff} (a). This kind of behavior has also been seen in contact interaction \cite{Wang:2022mrh} as well as lattice simulation result \cite{Alexandrou:2021ztx}. This may be due to the factor $x$ in the denominator of $F_S(Q^2)$. We have compared our scalar FF $F_S(Q^2)$ result with available lattice simulation \cite{Alexandrou:2021ztx} and contact interaction results \cite{Wang:2022mrh} in Fig. \ref{scalarff} (b), which is found to be in good agreement with them. Our result has a slightly higher distribution compared to both of them. At $Q^2=0$, the $F_S(Q^2)$ is found to be $1.323$ compared to $1.16$ for the case of contact interaction \cite{Wang:2022mrh}. In Fig. \ref{scalarff} (c), we have plotted the second moment of scalar FF along with the first moment. Overall, the scalar, vector, and tensor FFs calculated in our model are found to be in good agreement with available results. 

\begin{figure}[ht]
		\centering
		\begin{minipage}[c]{1\textwidth}\begin{center}
				(a)\includegraphics[width=.45\textwidth]{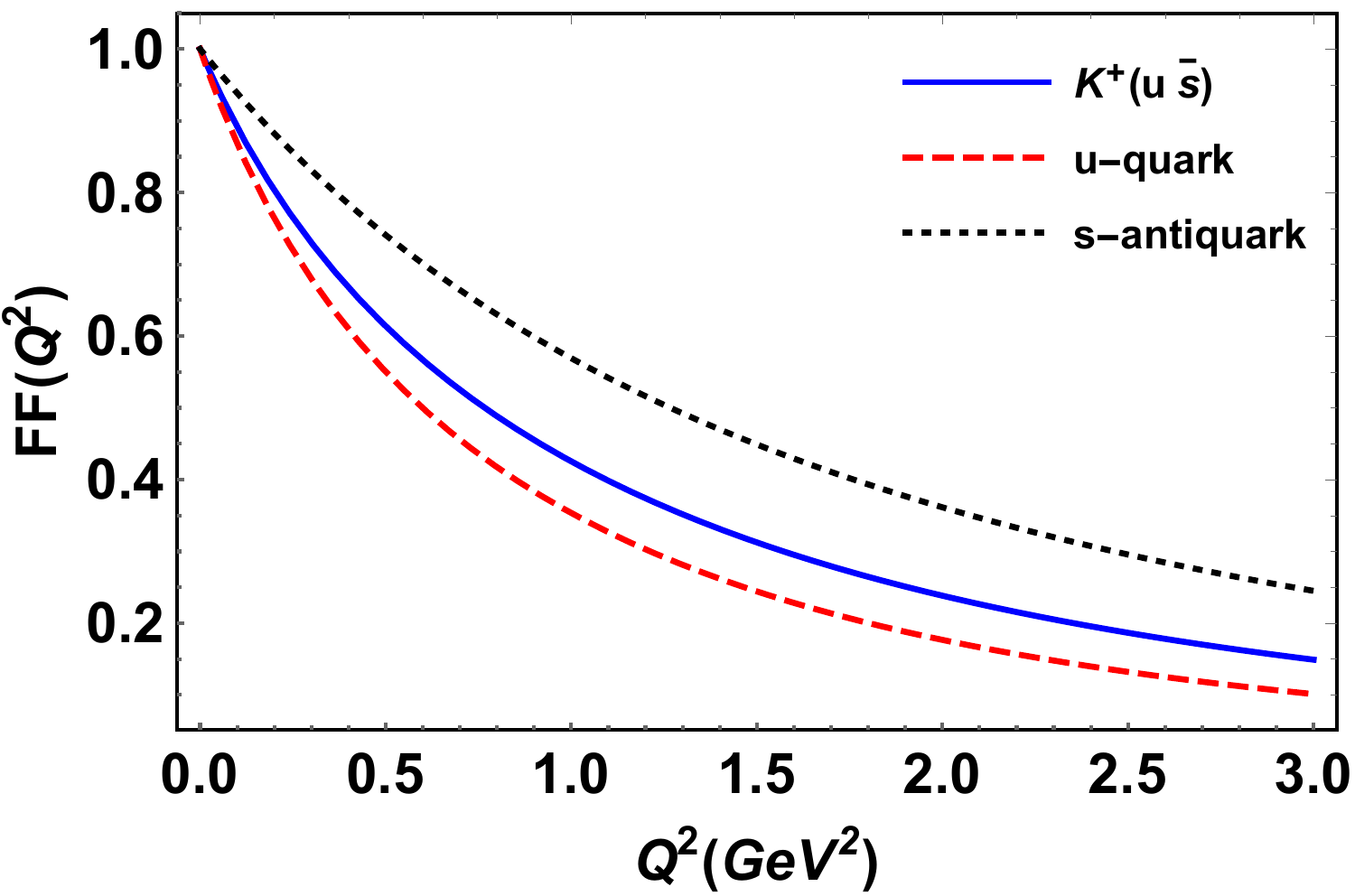}
				(b)\includegraphics[width=.45\textwidth]{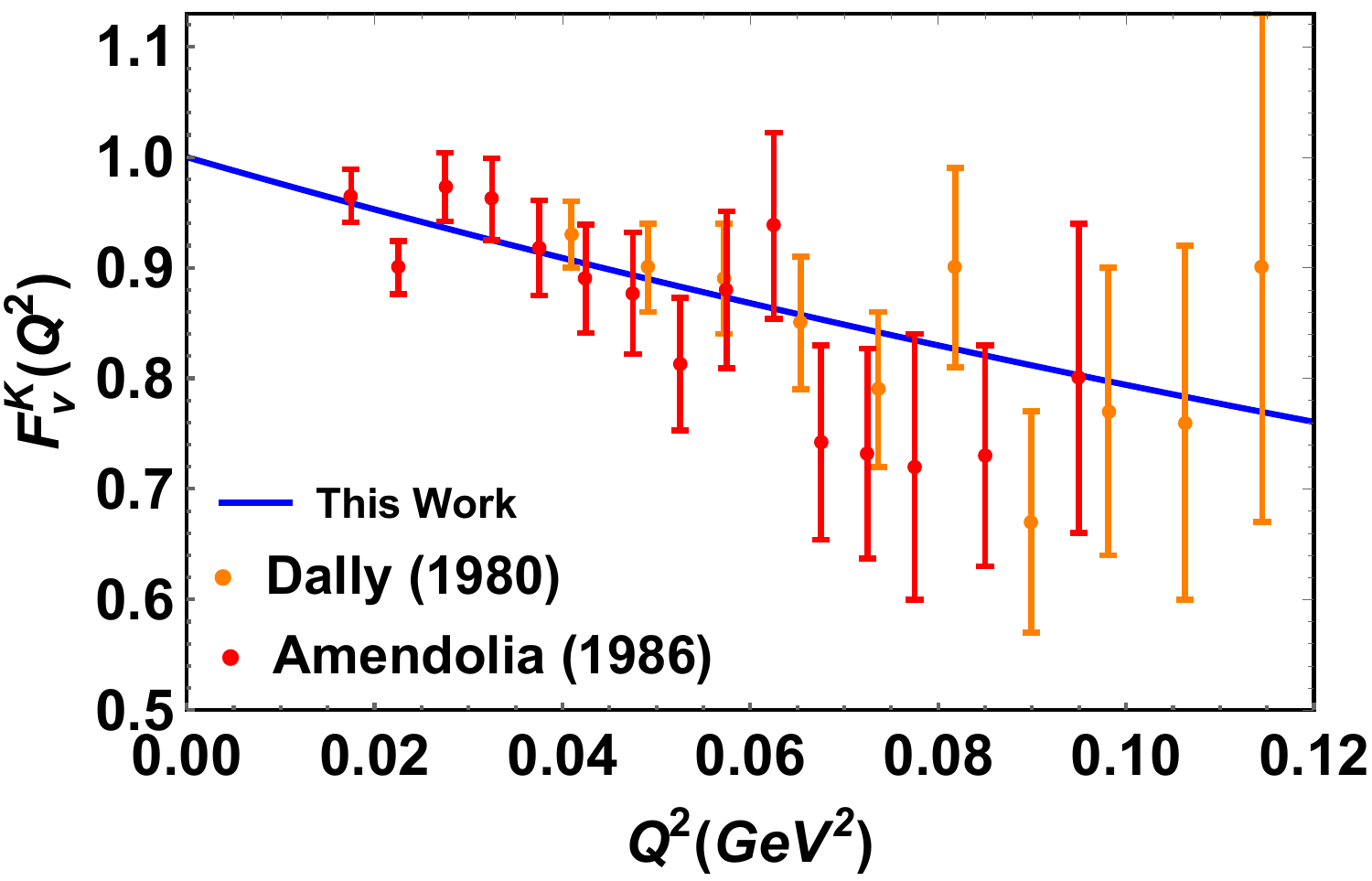}
			\end{center}
		\end{minipage}
		\begin{minipage}[c]{1\textwidth}\begin{center}
				(c)\includegraphics[width=.45\textwidth]{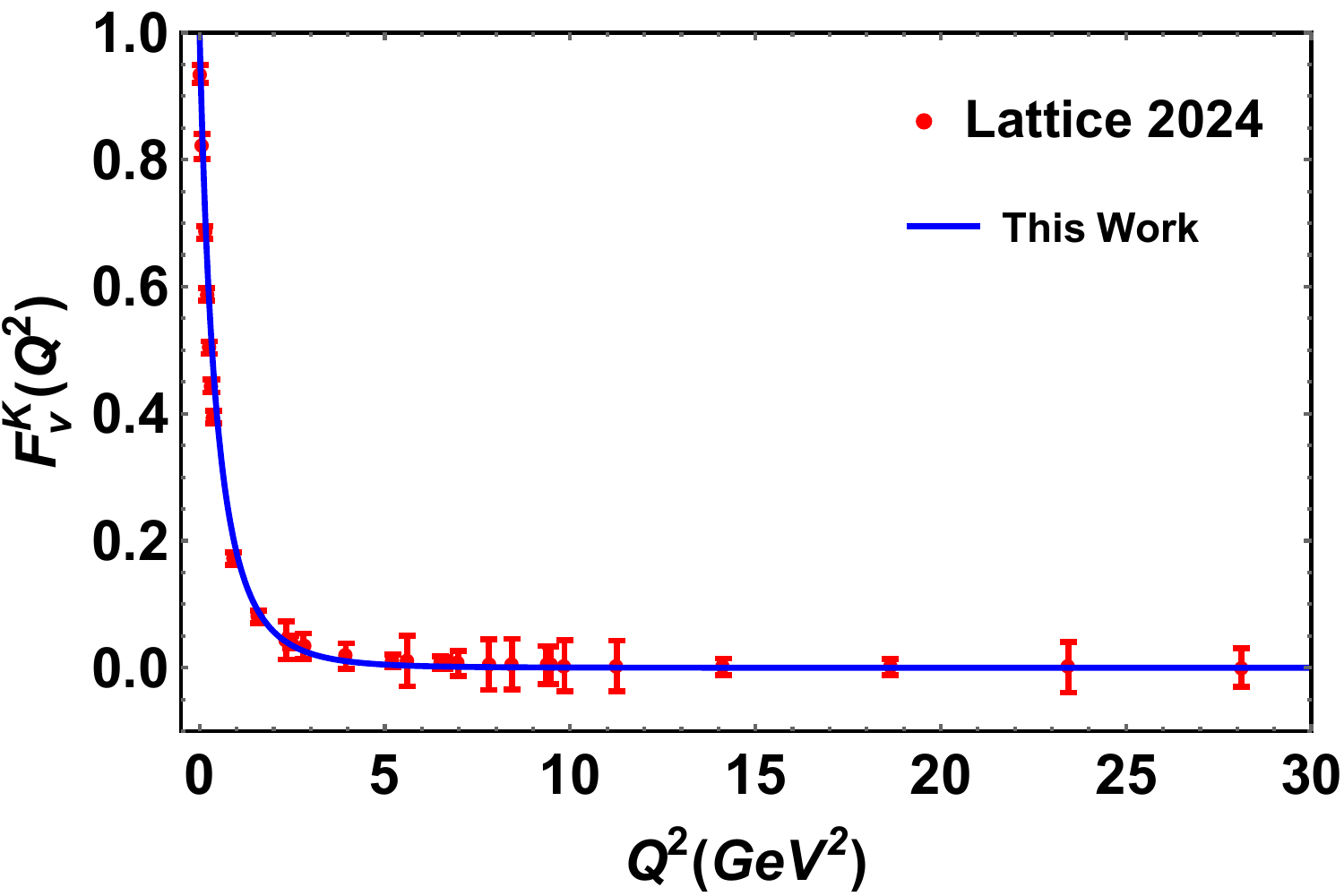}
                (d)\includegraphics[width=.45\textwidth]{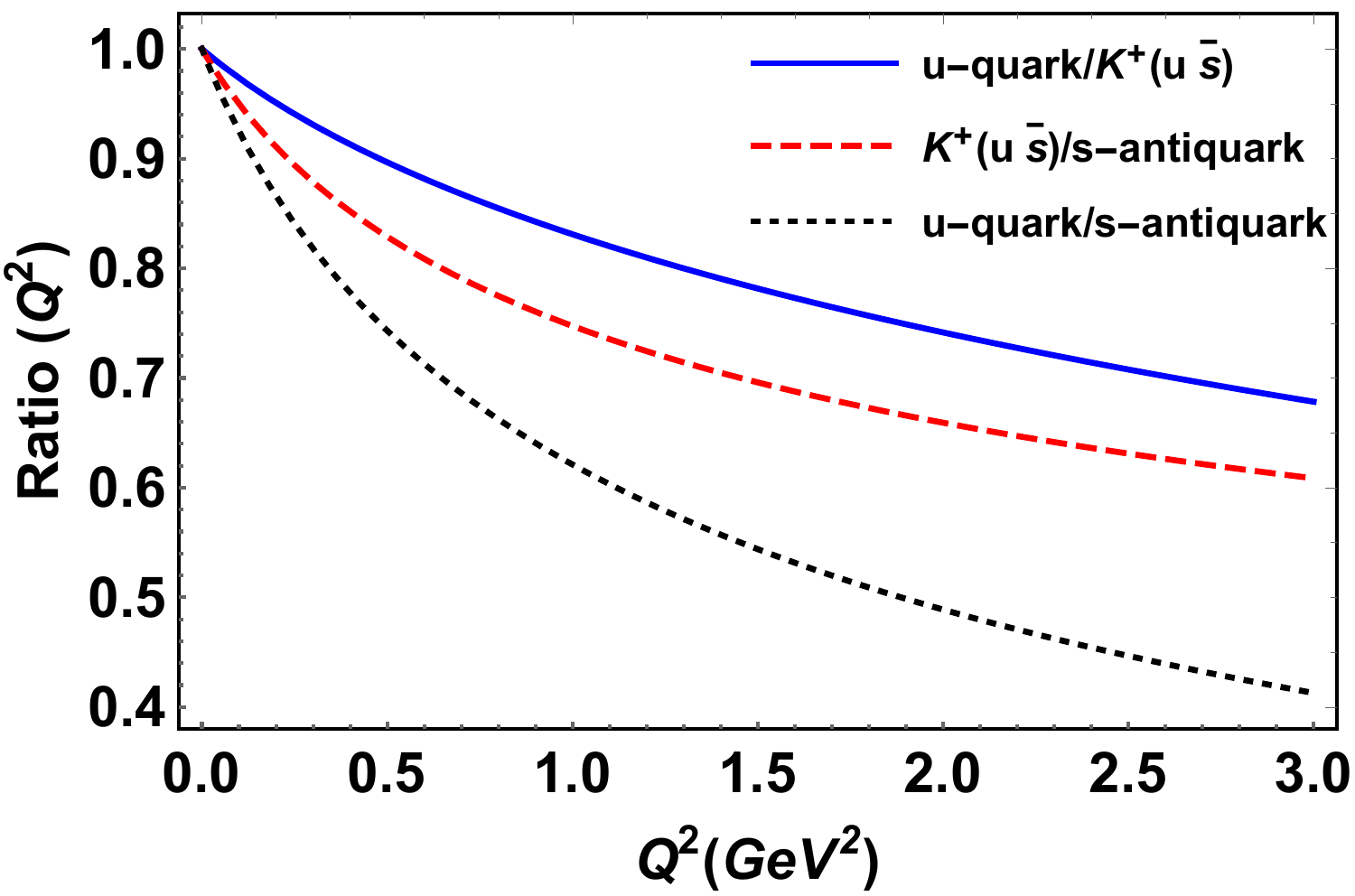}
                \end{center}
		\end{minipage}
		\caption{(Color online) The vector FF of kaon and its constituent ($u$-quark and $s$-antiquark) have been plotted with respect to $Q^2$ GeV$^2$. (a) The vector FF of $K^+(u \bar s)$, $u$-quark, and $s$-antiquark have been plotted with respect to $Q^2$ up to $3$ GeV$^2$. (b) The vector FF of kaon of our result has been compared with Dally \cite{Dally:1980dj} and Amendolia \cite{Amendolia:1986ui} experimental results up to $Q^2=0.12$ GeV$^2$. (c) The vector FF of kaon of our result has been compared with lattice simulation data \cite{Ding:2024lfj} up to $Q^2= 30$ GeV$^2$. (d) The $u$-quark/$K^+$, $K^+$/$\bar{s}$-antiquark, and $u$-quark/$\bar{s}$-antiquark FF ratios have been plotted with respect to $Q^2$ GeV$^2$.}
		\label{kaonvector}
	\end{figure}

\begin{figure}[ht]
		\centering
		\begin{minipage}[c]{1\textwidth}\begin{center}
				(a) \includegraphics[width=.45\textwidth]{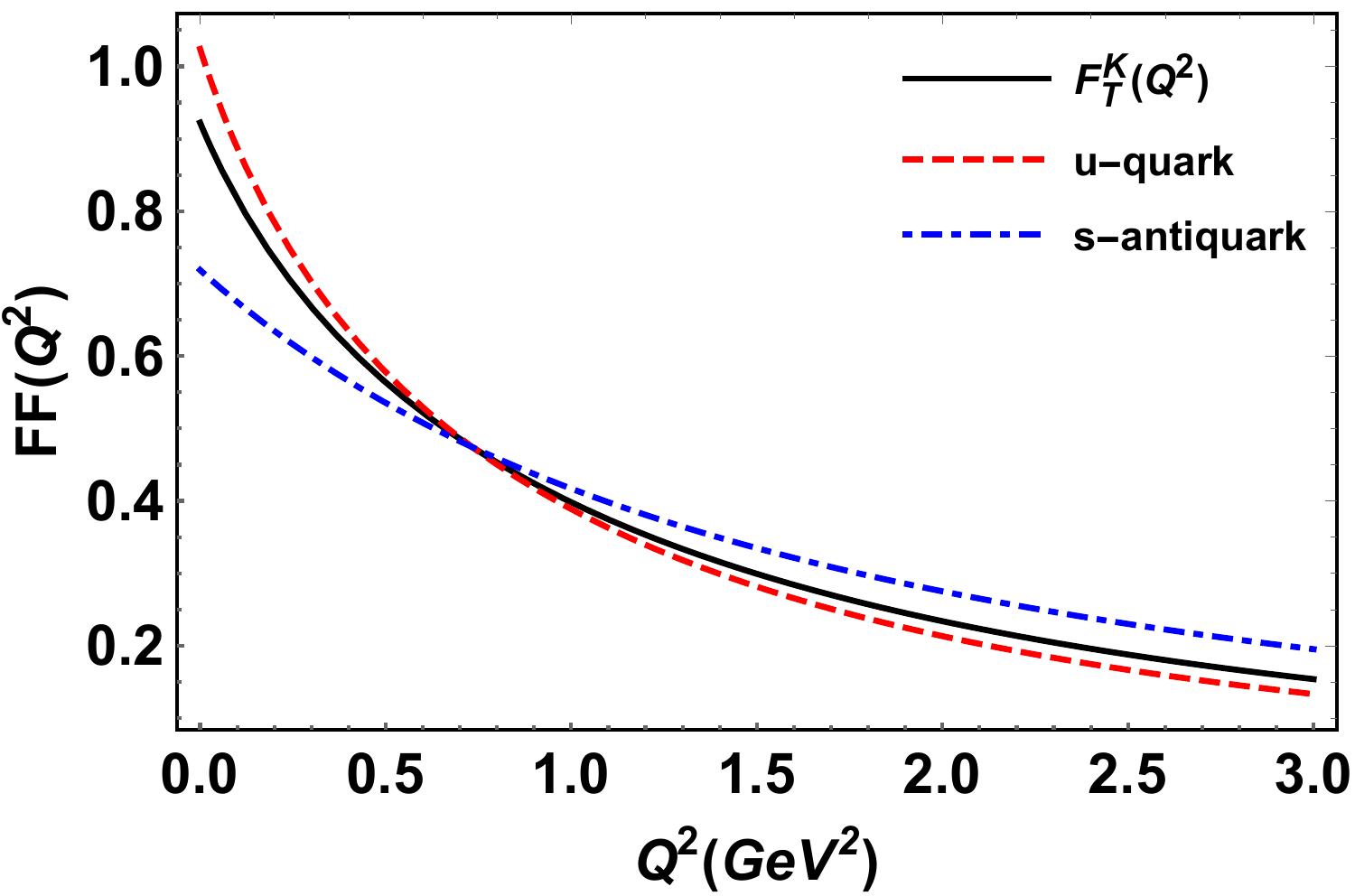}
				(b)\includegraphics[width=.45\textwidth]{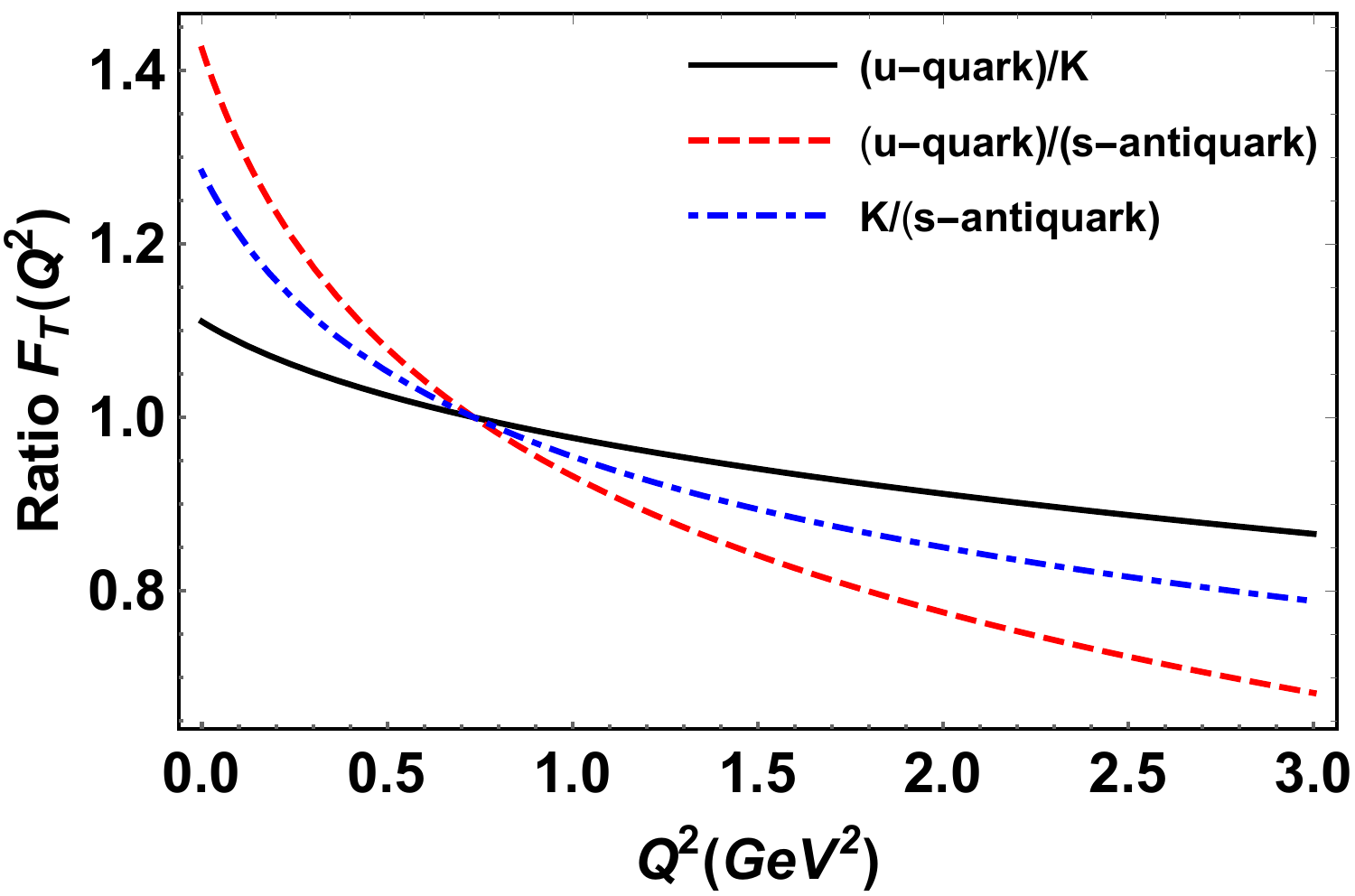}
			\end{center}
		\end{minipage}
		\begin{minipage}[c]{1\textwidth}\begin{center}
				(c)\includegraphics[width=.45\textwidth]{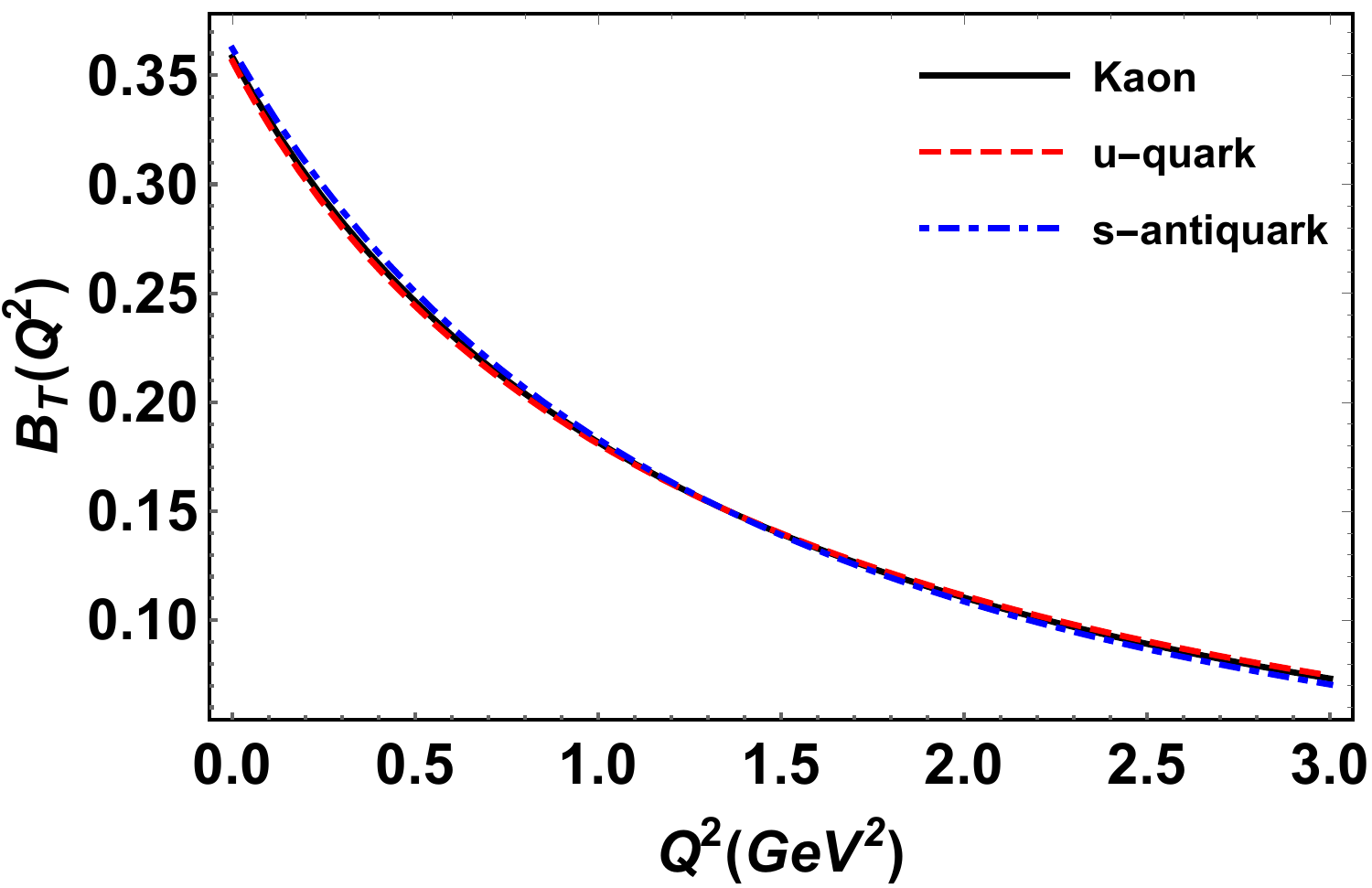}
                \end{center}
		\end{minipage}
		\caption{(Color online) (a) The tensor FF of kaon ($F_T^K(Q^2)$) and its constituents ($F_T^q(Q^2)$) (u-quark and s-antiquark) have been plotted with respect to $Q^2$ GeV$^2$. (b) The $u$-quark/$K^+$, $u$-quark/$\bar{s}$-antiquark, and $K^+$/$\bar{s}$-antiquark tensor FFs ratios have been plotted. (c) The second moment $B_T(Q^2)$ of the kaon and its constituents has been plotted.}
		\label{kaontensor}
	\end{figure}

\begin{figure}[ht]
		\centering
		\begin{minipage}[c]{1\textwidth}\begin{center}
				(a) \includegraphics[width=.45\textwidth]{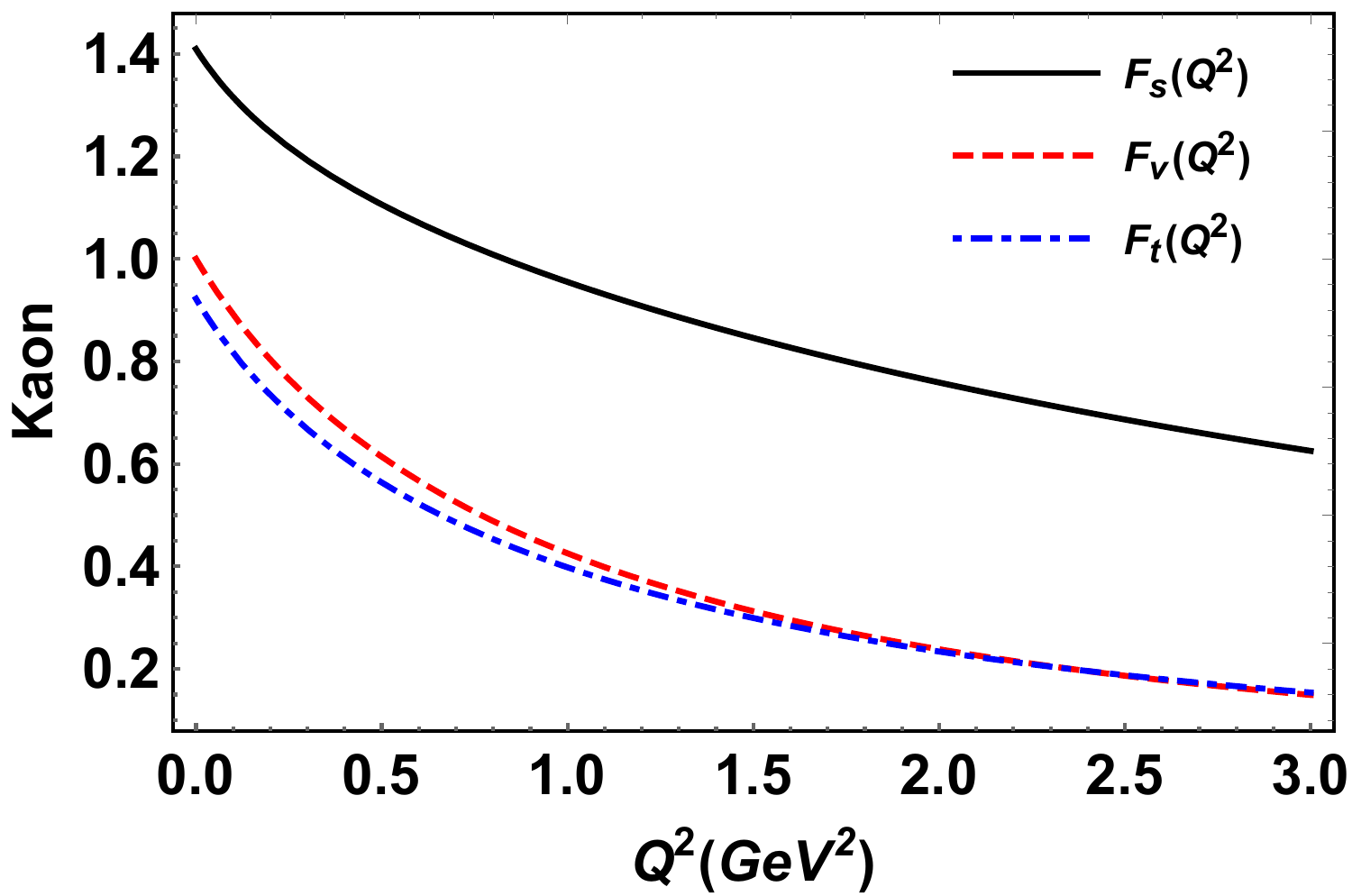}
				(b)\includegraphics[width=.45\textwidth]{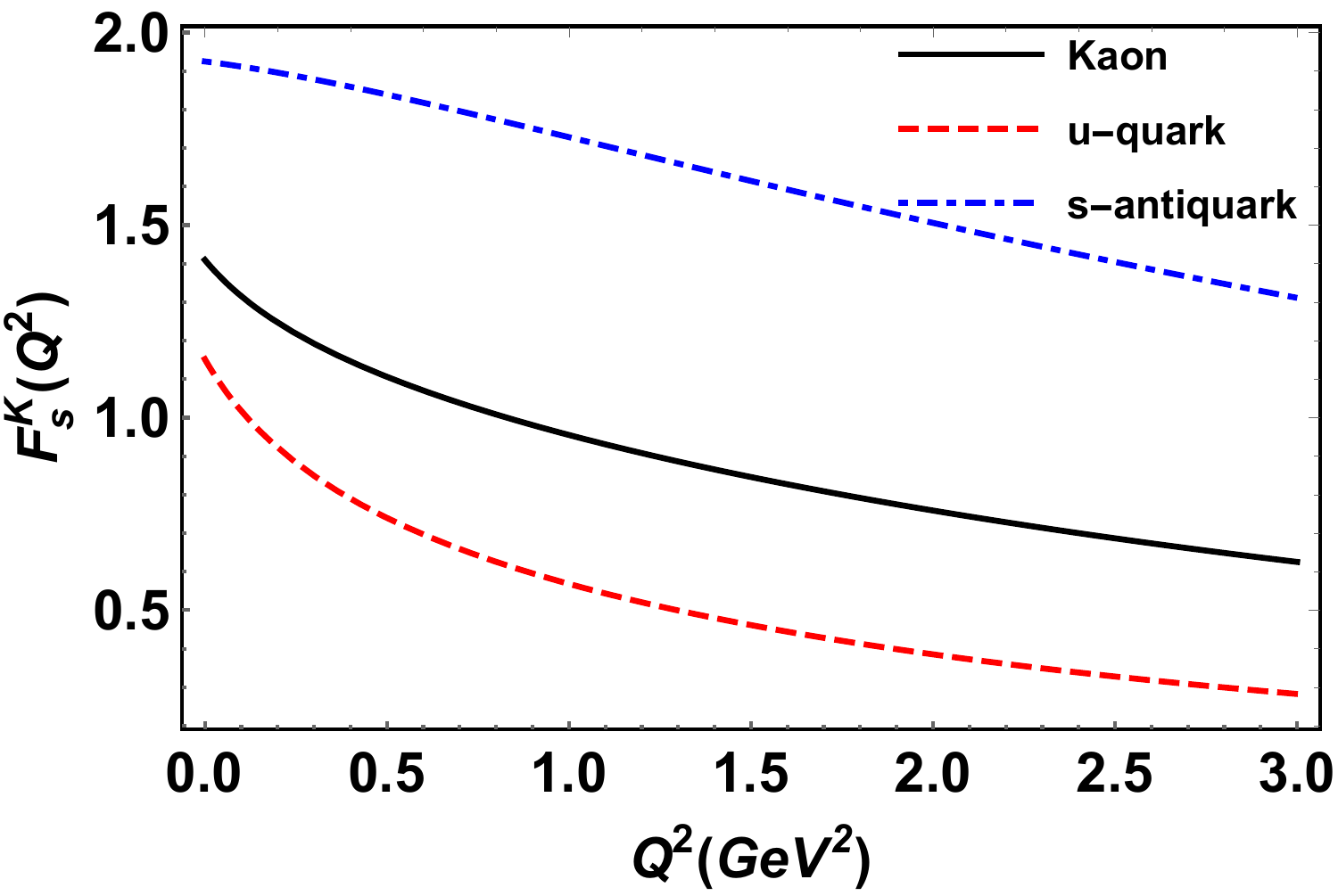}
			\end{center}
		\end{minipage}
		\begin{minipage}[c]{1\textwidth}\begin{center}
				(c)\includegraphics[width=.45\textwidth]{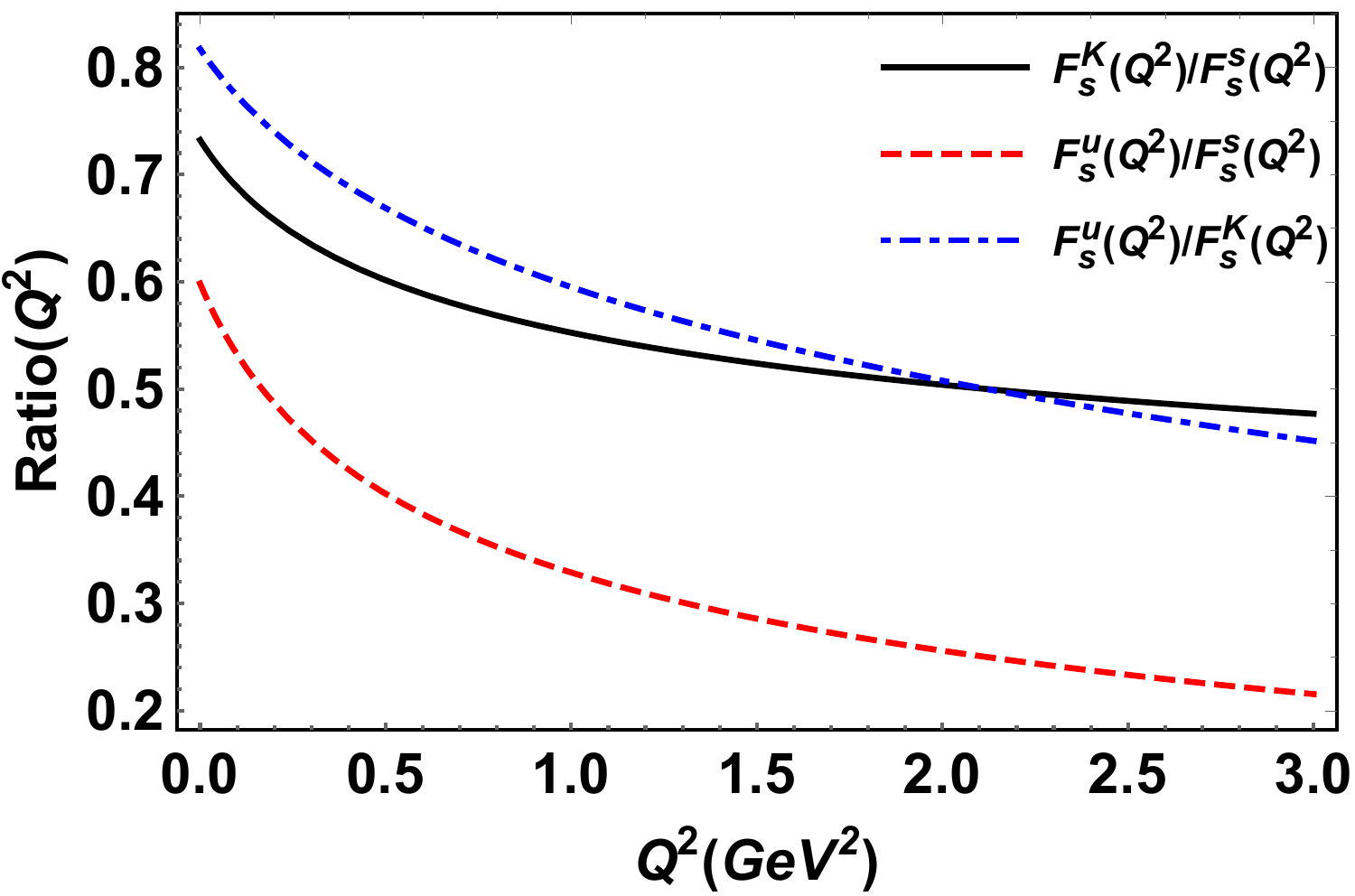}
                \end{center}
		\end{minipage}
		\caption{(Color online) (a) The scalar ($F_S(Q^2)$), vector ($F_v(Q^2)$), and tensor ($F_T(Q^2)$) FFs of kaon have been plotted with respect to $Q^2$ GeV$^2$. (b) The scalar FF of kaon ($F_S^{K^+} (Q^2)$) and its constituents ($F_S^{q} (Q^2)$) ($u$-quark and $\bar{s}$-antiquark) have been plotted. (c) The $K^+$/$\bar{s}$-antiquark, $u$-quark/$\bar{s}$-antiquark, and $u$-quark/$K^+$ scalar FF ratio have been plotted.}
		\label{kaonscalar}
	\end{figure}

\begin{figure}[ht]
		\centering
		\begin{minipage}[c]{1\textwidth}\begin{center}
				(a)\includegraphics[width=.45\textwidth]{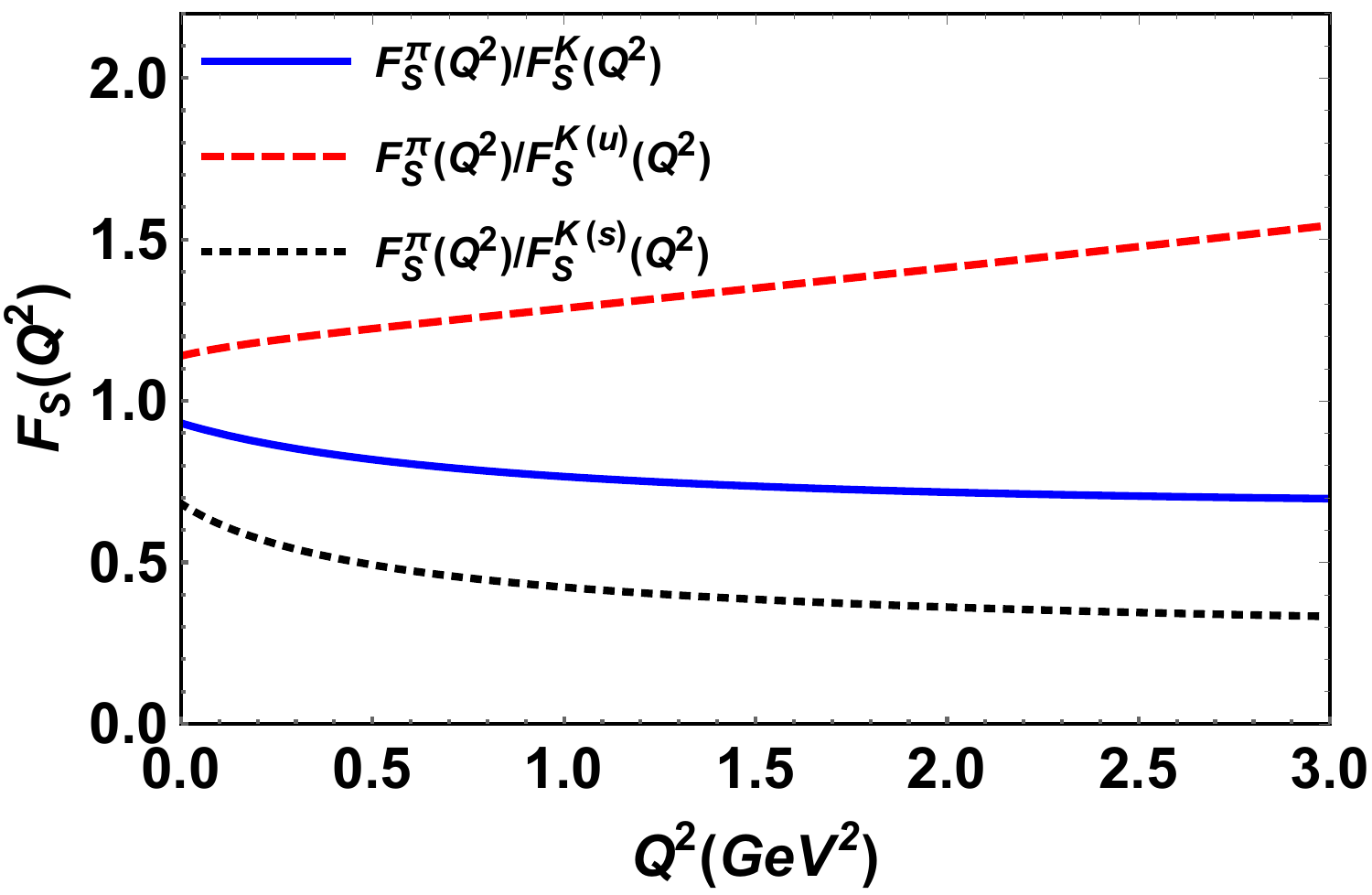}
				(b)\includegraphics[width=.45\textwidth]{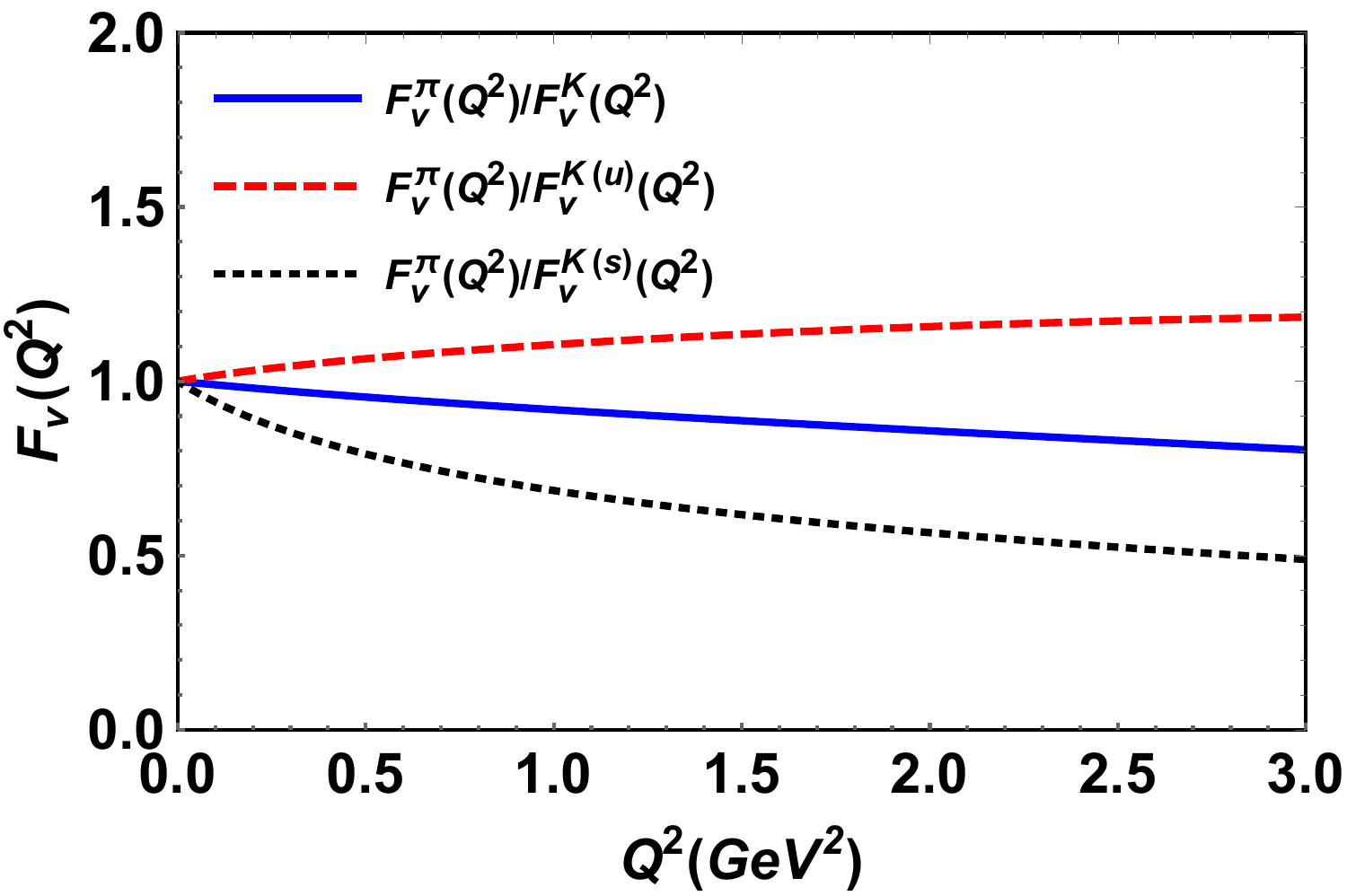}
			\end{center}
		\end{minipage}
		\begin{minipage}[c]{1\textwidth}\begin{center}
				(c)\includegraphics[width=.45\textwidth]{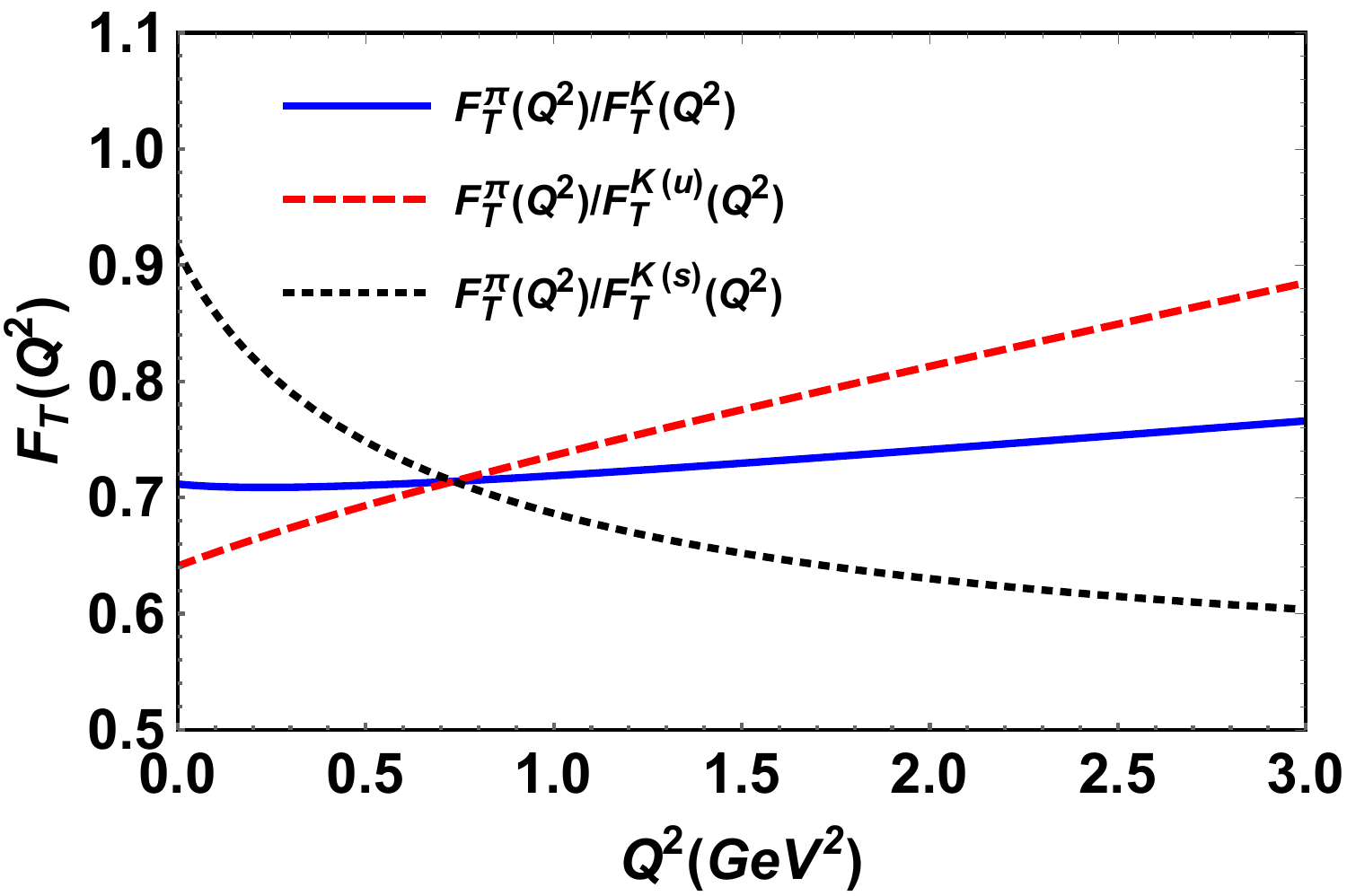}
                \end{center}
		\end{minipage}
		\caption{(Color online) 
      (a) The scalar, vector, and tensor FFs ratio of pion to kaon (along with $u$ quark and $\bar s$-antiquark ) have been plotted with respect to $Q^2$ in (a), (b), and (c), respectively.}
		\label{pionkaon}
	\end{figure}

Now moving into the results of FFs of kaon and its constituents. We have demonstrated the vector FFs of positively charged kaon, $u$-quark, and $\bar s$-antiquark of kaon in Fig. \ref{kaonvector}. The $\bar s$-antiquark is found to have a higher charge distribution compared to $u$-quark and kaon. This is due to the heavier mass of $\bar s$-antiquark than the $u$-quark. Similar kinds of observations have also been seen in Ref. \cite{Son:2024uet,Gao:2017mmp}. All the vector FFs show $F(Q^2=0)=1$ for the case of kaon. With the increase in $Q^2$, all the FFs saturate. In Fig. \ref{kaonvector} (b), we have plotted the vector FF $F_v^K(Q^2)$ of kaon up to $Q^2=0.12$ GeV$^2$ with available experimental results \cite{Dally:1980dj,Amendolia:1986ui} and found it to be in good agreement with both results. While for the higher $Q^2$ region, we have plotted the $F_v^K(Q^2)$ up to $Q^2=30$ GeV$^2$ of kaon with only available lattice simulation data \cite{Ding:2024lfj} and found it to exactly match with them. The vector FFs ratio of $F_v^u(Q^2)/F_v^K(Q^2)$, $F_v^K(Q^2)/F_v^{\bar s}(Q^2)$, and $F_v^u(Q^2)/F_v^s(Q^2)$ have been plotted in Fig. \ref{kaonvector} (d) and found out to decrease smoothly with $Q^2$. All these ratios are found to be unity at $Q^2=0$. It is observed that the ratio $F_v^u(Q^2)/F_v^K(Q^2)$ decreases slowly compared to that of others. The ratio $F_v^u(Q^2)/F_v^s(Q^2)$ of our result is found to follow a similar kind of trend when compared with Refs. \cite{Gao:2017mmp,Alexandrou:2021ztx}. The ratio of $F_v^s(Q^2)/F_v^u(Q^2)$ found to $3.45$ at $Q^2=6$ GeV$^2$, which is quite high compared to Ref. \cite{Gao:2017mmp}. This $F_v^u(Q^2)/F_v^s(Q^2)$ is found to decrease up to some positive value of $Q^2$ and then saturate at unity as $Q^2\rightarrow \infty $, as mentioned in Ref. \cite{Son:2024uet,Gao:2017mmp}. Now looking into the tensor FF of kaon, it is found that the $u$-quark has higher distributions compared to kaon and $\bar s$-antiquark in the region $0\le Q^2\le 0.73$ GeV$^2$ and vice versa after, $Q^2\ge0.75$ as shown in Fig. \ref{kaontensor}. There is no experimental data, and theoretical models have been reported for the case of kaon tensor and scalar FFs till now, but only lattice simulation data is available in Ref. \cite{Alexandrou:2021ztx}. In the case of Ref. \cite{Alexandrou:2021ztx}, it has been predicted that the tensor FF of both $u$-quark has a higher distribution than $\bar s$-antiquark of kaon, which can also be seen in our calculations. At $Q^2=0$, the tensor FF of $u$-quark, $\bar s$-antiquark, and kaon is found to be $1.02$, $0.71$, and $0.93$ respectively. In Fig. \ref{kaontensor} (b), we have shown the ratios $F_T^u(Q^2)/F_T^K(Q^2)$, $F_T^u(Q^2)/F_T^s(Q^2)$, and $F_T^K(Q^2)/F_T^s(Q^2)$ of kaon. The $F_T^u(Q^2)/F_T^s(Q^2)$ tensor FF ratio is found to follow the same trend as lattice simulations \cite{Alexandrou:2021ztx}. The second moment of the kaon tensor FF along with its constituent has been plotted in Fig. \ref{kaontensor} (c). All are found to have exactly similar distributions. The kaon scalar $F_S^{K}(Q^2)$, vector $F_v^{K}(Q^2)$, and tensor $F_T^{K}(Q^2)$ have been plotted in Fig. \ref{kaonscalar} (a). The scalar $F_S^{K}(Q^2)$ FF is found to have a higher distribution as compared to other FFs as the pion has a value of $1.42$ at $Q^2=0$. However, both vector $F_v^{K}(Q^2)$ and tensor $F_T^{K}(Q^2)$ have nearly similar behavior after $Q^2 \ge 1.75$. As like vector FFs results, the scalar $\bar s$-antiquark FF is found to have a higher distribution than $u$-quark and kaon, as shown in Fig. \ref{kaonscalar} (b). The scalar FFs ratios among the kaon and its constituents have been plotted in Fig. \ref{kaonscalar} and found to have similar behavior as Ref. \cite{Alexandrou:2021ztx}.
From both pion and kaon, we can conclude that all the FFs show 
\begin{eqnarray}
    F_S^{\pi(K)}(Q^2) \ge  F_v^{\pi(K)}(Q^2) \ge  F_T^{\pi(K)}(Q^2)
\end{eqnarray}
at low $Q^2 \le 1$ GeV$^2$ region. The pion to kaon, $u$-quark, and $\bar s$-antiquark FFs ratios have been shown in Fig. \ref{pionkaon}. The scalar FF $F_S^{\pi}(Q^2)/F_S^{u(K)}(Q^2$ ratio has contradictory behavior with lattice simulation results \cite{Alexandrou:2021ztx}, while others have a similar kind of trend with lattice data.
\par We have also calculated the radii corresponding to different FFs for both pion and kaon. The scalar, vector, and tensor radii can be calculated as \cite{Wang:2022mrh}
\begin{eqnarray}
    \langle r_S^2\rangle &= & \frac{-6}{F_S(0)} \frac{\partial F_S(Q^2)}{\partial Q^2}\Big|_{Q^2\rightarrow0}, \\
    \langle r_V^2\rangle &= & \frac{-6}{F_V(0)} \frac{\partial F_V(Q^2)}{\partial Q^2}\Big|_{Q^2\rightarrow0},\\
    \langle r_T^2\rangle &= & \frac{-6}{F_T(0)} \frac{\partial F_T(Q^2)}{\partial Q^2}\Big|_{Q^2\rightarrow0}.
\end{eqnarray}
\begin{table*}
\begin{tabular}{|c|c|c|c|c|c|c|c|c|c|c|c|c|}
\hline  & \multicolumn{3}{c|}{Pion} & \multicolumn{3}{c|}{u-quark (Kaon)}& \multicolumn{3}{c|}{s-antiquark (K)}& \multicolumn{3}{c|}{Kaon}\\
\cline{2-13}
 & $\sqrt{\langle r^2_S \rangle}$  & $\sqrt{\langle r^2_V \rangle}$ & $\sqrt{\langle r^2_T \rangle}$ &  $\sqrt{\langle r^2_S \rangle}$ & $\sqrt{\langle r^2_V \rangle}$ & $\sqrt{\langle r^2_T \rangle}$ &$\sqrt{\langle r^2_S \rangle}$ & $\sqrt{\langle r^2_V \rangle}$ & $\sqrt{\langle r^2_T \rangle}$ & $\sqrt{\langle r^2_S \rangle}$ & $\sqrt{\langle r^2_V \rangle}$ & $\sqrt{\langle r^2_T \rangle}$ \\
\hline
This Work & 0.528 & 0.558 & 0.567 & 0.440 & 0.663 & 0.605 & 0.338 & 0.406 & 0.326 & 0.409 & 0.568 & 0.529\\
\hline
LQCD \cite{Alexandrou:2021ztx} & 0.482 & 0.539 & 0.679 & 0.386 &- & 0.618 & 0.321 & - & 0.5 & - & 0.538 &- \\
\hline
CI-MRL \cite{Wang:2022mrh} & 0.434 & 0.558 & 0.583 &- &- &-&-&-&-&-&-&- \\
\hline
Ref. \cite{Cui:2021aee}  & - & 0.640 & - &- &- &-&-&-&-&-&0.530&- \\
\hline
NA7 \cite{Amendolia:1984nz,NA7:1986vav}& - & 0.657, 0.662 & - &- &- &-&-&-&-&-&-&- \\
\hline
JLQCD \cite{Aoki:2015pba}  & - & 0.677 & - &- &- &-&-&-&-&-&0.616&- \\
\hline
CERN-NA007 \cite{Amendolia:1986ui}  & - & - & - &- &- &-&-&-&-&-&0.583&- \\
\hline
PDG \cite{ParticleDataGroup:2020ssz}  & - & 0.659 & - &- &- &-&-&-&-&-&0.560&- \\
\hline
LQCD \cite{Gulpers:2013uca} & 0.635 & - & - &- &- &-&-&-&-&-&-&- \\
\hline
Ref. \cite{Gao:2017mmp} & - & 0.65 & - & - & 0.62 &-&-&0.43&-&-&-&- \\
\hline

\end{tabular}
\caption{The scalar, vector, and tensor radii of pion, kaon, and their constituents in unit of fermi (fm) in our results have been compared with availabe experimental results \cite{Amendolia:1984nz,NA7:1986vav,Amendolia:1986ui,ParticleDataGroup:2020ssz}, lattice QCD simulation results \cite{Alexandrou:2021ztx,Aoki:2015pba,Gulpers:2013uca}, and theoretical models \cite{Wang:2022mrh,Cui:2021aee,Gao:2017mmp}.}
\label{tableradii}
\end{table*}
In our case, $F_V(0)$ is found to be unity for both the pion and kaon. The radii for $u$-quark and $\bar s$-antiquark are calculated from quark FFs. The predicted values of pion, $u$-quark of kaon, $\bar s$-antiquark, and charged kaon have been presented in Table \ref{tableradii}. We have also compared our results with available experimental data \cite{Amendolia:1984nz,NA7:1986vav,Amendolia:1986ui,ParticleDataGroup:2020ssz}, lattice simulations \cite{Alexandrou:2021ztx,Aoki:2015pba,Gulpers:2013uca}, and theoretical predictions \cite{Wang:2022mrh,Cui:2021aee,Gao:2017mmp}. For the case of the pion, we found that the radii follow the order
\begin{eqnarray}
    \sqrt{\langle r_T^2\rangle} (fm) \ge \sqrt{\langle r_V^2\rangle} (fm) \ge \sqrt{\langle r_S^2\rangle} (fm). 
\end{eqnarray}
Similar behavioral orders are reported in Ref. \cite{Wang:2022mrh,Alexandrou:2021ztx}. While for the case of $u$-quark of kaon, $\bar s$-antiquark of kaon, and charged kaon, the order is found to be  
\begin{eqnarray}
     \sqrt{\langle r_V^2\rangle} (fm) \ge \sqrt{\langle r_T^2\rangle} (fm) \ge \sqrt{\langle r_S^2\rangle} (fm).
\end{eqnarray}
However, in Ref. \cite{Alexandrou:2021ztx}, the tensor radii have a higher value than the vector radii, which contradicts our case. The $\sqrt{\langle r_V^s \rangle}/\sqrt{\langle r_V^{u}\rangle}$ vector charge radii ratio of kaon is found to be $0.61$ compared to $0.69$ of Ref. \cite{Gao:2017mmp}. As there is no data available for the case of scalar and tensor radii of kaon, so we can not compare them. However, the overall radii are found to be in good agreement with other results. Upcoming EIC and JLab will provide more insight into the vector charge radii of the pion and kaon.

\section{Conclusion}\label{con}
In this work, we have investigated the scalar, vector, and tensor form factors (FFs) of the pion and kaon within the light-cone quark model framework. These FFs have been derived from the leading and sub-leading twist generalized parton distribution functions. By solving the quark-quark correlator, we have presented the FFs in the form of light-front wave functions and calculated their explicit forms by using the proper spin wave functions. We have demonstrated these FFs through two-dimensional plots. 
We have compared our results with available experimental data, lattice simulation results, and available theoretical predictions. We have also calculated the second moment of these FFs and evolved them to a higher scale to compare with the available lattice simulation results. All our results are found to be in good agreement with other data. The vector FFs of the pion and kaon obey the $F_v(Q^2=0)=1$ property. We observed that the scalar FFs have higher distributions compared to the vector and tensor FFs for both pion and kaon. We have also studied the FFs of $u$-quark and $\bar s$-antiquark of kaon. The $\bar s$-antiquark shows a higher distribution compared to $u$-quark for both the vector and scalar FFs, while the opposite behavior was seen for the case of the tensor FF. There is no work reported in any theoretical models for tensor and scalar FFs yet. The kaon results were found to have the same trend as lattice simulation results, but the magnitude differs from them. We have also predicted the scalar, vector, and tensor radii of pion and kaon (along with its constituents), calculated from their respective FFs. The radii are found to be in good agreement with available experimental results, lattice simulation data, and theoretical model predictions. 
\par There is still no experimental data available for scalar and tensor FFs, however, some data are available for the vector FFs case for pion and kaon. Even though there is limited studies going on for the case of kaon FFs in experiment and lattice simulation, this study will definitely provide deeper insights into the importance of tensor and scalar contributions of pseudoscalar mesons in understanding the internal structure of hadrons.


\section{Acknowledgement}
H.D. would like to thank  the Science and Engineering Research Board, Anusandhan-National Research Foundation, Government of India under the scheme SERB-POWER Fellowship (Ref No. SPF/2023/000116) for financial support.

\section{Reference}

\bibliography{ref}

\end{document}